\documentclass[prb,twocolumn,showpacs,superscriptaddress,amsmath,amsfont,floatfix]{revtex4}
\usepackage{graphicx,graphics}
\usepackage{amssymb}
\usepackage{natbib}
\usepackage{citesort}
\usepackage{color}
\usepackage{ulem}
\usepackage{subfigure}

\begin{document}

\title{Normal state of metal-intercalated phenacene crystals: Role of electron correlations}
\author{Tirthankar Dutta} 
\author{Sumit Mazumdar}
\affiliation{Department of Physics, University of Arizona, Tucson, AZ 85721} \date{\today}
\begin{abstract}
In this work we study the effect of long range electron-electron correlations on the behavior of the normal state 
of metal-intercalated phenacene crystals. While the individual phenacene molecules are modeled by the Pariser-Parr-Pople 
Hamiltonian with long range Coulomb interactions, we derive a correlated minimal model for describing the phenacene ionic 
crystals. We find that long range electron correlations do not change the behavior of the phenacene ions with molecular valence 
$-1$ (monoanion) and $-2$ (dianion), compared to that observed for short range electron interactions. The monoanion crystal
is a single-band $\frac{1}{2}$-filled antiferromagnetic Mott-Hubbard semiconductor while the dianion crystal is a two-band 
semiconductor with inter-molecular antiferromagnetic and intra-molecular ferromagnetic spin orderings. However, the trianion 
crystal is no longer a nearly degenerate $\frac{3}{4}$-filled two-band system. We show that this occurs because the kinetic 
stability of the $\frac{3}{4}$-filled two-band system with long range correlations is smaller compared to that with short range 
correlations, for the lattice sizes considered in this study. We argue that with large finite lattices, behavior of the trianion 
crystal with long range correlations will be same as that with short range correlations. We thus conclude that long range 
correlations fail to alter the normal states of metal-intercalated phenacene crystals.   
\end{abstract}
\maketitle

\section{Introduction}

Strongly correlated multi-orbital systems lie at the forefront of modern day condensed matter physics. Interaction between the 
charge, spin and orbital degrees of freedom of the electrons in these systems leads to various novel phenomena 
like, giant magnetoresistance, itinerant ferromagnetism, charge-orbital ordering, and superconductivity (SC) 
\cite{Tokura00a,Imada98a,Mackenzie03a,Nakatsuji00a,Khaliullin00a,Khaliullin01a,Ishihara02a,Maeno94a}. 
Among the multi-orbital superconductors, many show SC at a fixed or very narrow region of carrier 
density. The ones which are of particular interest to us are the carbon(C)-based unconventional superconductors such as the 
charge-transfer solids (CTS) \cite{Ishiguro}, alkali metal-doped fullerides ($A_{3}C_{60}$) \cite{Gunnarsson97a,Iwasa03a,Capone09a}, 
and the metal-intercalated phenacenes
\cite{Mitsuhashi10a,Kubozono11a,Wang11a,Wang11b,Wang12a,Artioli14a,Yano14a}. Experimentally it is found that SC in the CTS occur 
under pressure and at the carrier density $\rho \approx$ 0.5 \cite{Merino01a,Clay12b}; $\rho$ is defined as the ratio of the 
number of charge carriers to the number of active orbitals, i.e., $N_{e(h)}/N_{o}$, where $e$($h$) refer to electron(hole).
Similarly, doped fullerides and metal-intercalated phenacenes are found to show SC under pressure and at molecular valence $-3$, 
which correspond to $\rho =$ 1.5 \cite{Dutta14a}; note that $\rho =$ 1.5 (electrons as charge carriers) and $\rho =$ 0.5 (holes as 
charge carriers) are equivalent. Three different classes of C-based materials showing SC at a fixed carrier 
concentration raises an important question: Is there something special about carrier density 0.5? 
\\
\indent 
In order to understand the mechanism of SC in these multi-orbital C-based systems it is important to first understand the normal 
state at  this particular carrier concentration. It is with this motivation that we began the investigation of the electronic 
structure of metal-doped phenacenes \cite{Dutta14a}, which have been claimed to be superconducting
\cite{Mitsuhashi10a,Kubozono11a,Wang11a,Wang11b,Wang12a,Artioli14a,Yano14a} only at molecular valence $-$3. After the discovery 
of SC in metal-intercalated phenacenes, other groups have failed to reproduce the results obtained by Kubozono {\it et. al.} for 
picene \cite{Mitsuhashi10a,Kubozono11a} and Wang {\it et. al.} for phenanthrene \cite{Wang11a,Wang11b,Wang12a}. However recently, 
Artioli {\it et. al.} have reported SC in Sm-doped phenanthrene, chrysene, and picene \cite{Artioli15a}. They conclude that the
presence of SC at molecular valence $-3$ irrespective of the number of phenacene rings ``raises new questions that open the way for 
further and new theoretical investigations on the electron pairing mechanism'' in polycyclic aromatic hydrocarbons. 
\\
\indent
In addition to the work of Artioli {\it et. al.}, we believe that a proper theoretical understanding of the normal 
state of metal-intercalated phenacenes (phenanthrene and picene) is of interest from another perspective, namely, understanding
pressure-induced antiferromagnetic-to-SC transition in A15 Cs$_{3}$C$_{60}$ \cite{Ganin08a,Ganin10a}. In what follows, the terms 
monoanion, dianion and trianion will refer to $M^{1-}$, $M^{2-}$ and $M^{3-}$, respectively; here, $M$ refer to a molecule and the 
superscript denote the charge of $M$, i.e., number of doped electrons. In both phenacenes and fullerides (for which experimental 
results are robust), there is apparent breaking of charge conjugation symmetry, i.e., crystals of mononions and trianions behave 
differently. While molecular solids of phenacene and fulleride trianions show SC, crystals of their monoanions are semiconducting. 
In neutral $C_{60}$ molecules, the Lowest Unoccupied Molecular Orbitals (LUMOs) are empty and triply degenerate; the levels are 
referred to as $t_{1u}$. Upon electron doping, by virtue of Jahn-Teller distortion and electron-electron (e-e) interactions, this 
degeneracy is lifted; for illustration sake we shall denote these non-degenerate levels as $t^{n}_{1u}$, where the superscript $n$
refer to the orbital occupancy. Thus, each $C^{3-}_{60}$ ion consists of a low lying $t^{2}_{1u}$ orbital, a $t^{1}_{1u}$ orbital 
at intermediate energy, and a high lying $t^{0}_{1u}$ orbital. Similarly, each $C^{1-}_{60}$ ion have a low lying $t^{1}_{1u}$ 
orbital and two degenerate $t^{0}_{1u}$ orbitals. It is widely believed that the ambient pressure antiferromagnetism in 
$C_{60}^{3-}$ crystals arise from inter-molecular spin-spin coupling between electrons occupying the $t^{1}_{1u}$ orbitals. This 
$\frac{1}{2}$-filled Mott insulator under pressure becomes superconducting due to increased band-width 
\cite{Takabayashi09a,Capone09a,Capone02a,Nomura16a}. This viewpoint of SC in the fullerides however has some drawbacks. 
Firstly, this theory is based on the results of mean field and dynamical mean field theory (DMFT) studies showing 
antiferromagnetic-to-SC transition in the frustrated $\frac{1}{2}$-filled Hubbard model 
\cite{Schmalian98a,Vojta99a,Kyung06a,Powell05a,Powell07a}. However, more recent and sophisticated numerical calculations which go
beyond mean field level have shown the absence of SC in the frustrated $\frac{1}{2}$-filled Hubbard model 
\cite{Clay08a,Dayal12a,Tocchio09a,Watanabe08a,Gomes13a,Yanagisawa13a}. Secondly, this approach fails to explain the absence of SC in 
fulleride monoanion crystals which are also $\frac{1}{2}$-filled antiferromagnetic Mott insulators; the $t^{1}_{1u}$ orbitals of 
$C_{6}^{1-}$ ions can couple antiferromagnetically in the solid state. Lastly, the absence of SC in the Mott-Hubbard semiconductor 
K$_{1}$pentacene \cite{Craciun09a} confirms the theories that predict absence of SC within the $\frac{1}{2}$-filled band Hubbard 
model. Thus, a proper understanding of the normal state and hence the mechanism of SC in the fullerides is still not complete.  
\\
\indent 
In our previous study on phenancenes \cite{Dutta14a} we have shown that breaking of charge conjugation symmetry is a band-width 
induced effect, that gets further enhanced by electron correlations. A similar scenario is expected to hold for doped fullerides 
also, thereby leading to an alternative description of SC in doped fullerides within our theory of metal-intercalated phenacenes. 
At ambient pressure molecular solids of both $C^{1-}_{60}$ and $C_{60}^{3-}$ are $\rho =$ 1 antiferromagnetic Mott insulators. Under 
pressure there is equalization of electron population between the $t^{2}_{1u}$ and $t^{1}_{1u}$ levels of the trianions by virtue of 
inter-ion inter-orbital hoppings leading to the formation of a $\rho =$ 3/2 two-band paramagnetic system. SC in this case will be 
related to populations of 3/2 in two non-degenerate levels which become close in energy by virtue of increased band-width, and the 
pairing is inter-molecular rather than intra-molecular. Such a band-width induced population exchange will not occur in the 
monoanion crystal by virtue of its molecular charge and it continues to be a $\rho =$ 1 antiferromagnet (AFM). We are unaware of 
any experiment in this context that would preclude our proposed mechanism.
\\
\indent  
In our previous work \cite{Dutta14a}, we have explicitly shown that in the presence of short range e-e interactions, the normal 
states of the monoanion and trianion crystals are not identical but differs substantially. These different normal states under 
pressure would have different behaviors and thus, while the monoanion crystal is an AFM, the trianion solid shows SC. A summary of 
our previous results is presented in Sec-III. The motivation behind the present study is to understand the effect of long range e-e 
interactions on the normal states of the phenacene ionic crystals. Our modeling of charged phenacene crystals begins from an 
atomistic description of the individual molecules, described within a $\pi$-electron model. We construct localized molecular 
orbitals (MOs) for the individual phenacene molecules from which the description of the crystal is built using the frontier MOs 
(FMOs) as basis orbitals. Previously we had addressed the normal state of metal-intercalated phenanthrenes by deriving an effective 
low-energy Hamiltonian, $H^{Hub}_{L,L+1}$, within the localized basis of the LUMO (L) and LUMO+1 (L+1) FMOs \cite{Dutta14a}, and 
describing the individual phenanthrene molecules by the Hubbard model having only onsite (short range) e-e correlations. But 
C-based materials like the phenancenes are known to possess long range Coulomb correlations and thus, the Hubbard model is 
inadequate for their complete description. In the present work we model the individual phenanthrene molecules by the 
Pariser-Parr-Pople Hamiltonian (PPP) \cite{PPP53a,PPP53b} having true long range e-e interactions. The effective crystal Hamiltonian 
in this case is depicted by $H^{PPP}_{L,L+1}$. We find that long range interactions, apart from renormalizing the individual terms 
in $H^{Hub}_{L,L+1}$, introduce an additional repulsion term between {\it same spin} electrons occupying different orbitals on the 
same molecule (see Sec-II). We thus anticipate that long range correlations will not affect the normal state of the monoanion 
crystal but should alter the behavior of the dianion and trianion crystals. We find that long range e-e interactions only alter the 
behavior of the trianion crystal, although the monoanion-trianion asymmetry {\it persists}. We will show in Sec-IV that the change 
in behavior of the trianion crystal is due to the size of the lattices that we have considered. These small system sizes fail to 
provide the kinetic stability needed by the trianion crystal to behave as a nearly degenerate two-band $\frac{3}{4}$-filled system 
(see Sec-III). 

\section{Theoretical Modeling and Computational Details}

\begin{figure}[tb]
\includegraphics[width=3.0in,height=2.0in]{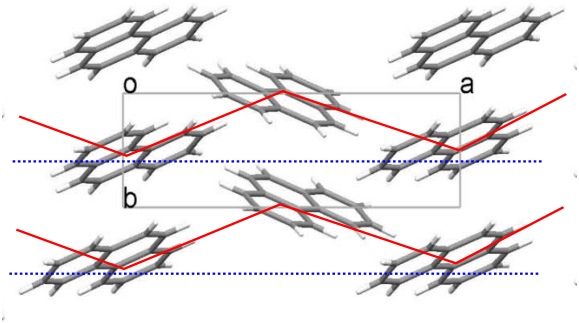}
\caption{(Color online) 2D Herringbone lattice of phenanthrene ions (the crystallographic motif is taken from 
Ref.~\onlinecite{IUCR}); metal ions are not shown for clarity. The 1D lattice studied in text is shown superimposed on the 2D 
lattice by solid red and blue dashed lines, depicting nearest neighbor and next-nearest neighbor hoppings, respectively.} 
\label{phlat}
\end{figure}

Metal-intercalated phenacenes crystallize in the Herringbone motif which is shown in Fig.~\ref{phlat}. The complete Hamiltonian for 
such a solid can be be written as $H$ = $H_{intra}$ + $H_{inter}$, where $H_{intra}$ depict the intra-molecular Hamiltonian of 
individual phenacene molecules and 
$H_{inter}$ represent the inter-molecular interactions. The intra-molecular Hamiltonian can be expressed as, 
\begin{eqnarray}
H_{intra} &=&-\epsilon \sum_{\mu,i}{}^{'} n_{\mu,i} - t \sum_{\mu,\langle ij \rangle,\sigma} c_{\mu,i,\sigma}^{\dagger}c_{\mu,j,\sigma}  + \nonumber \\
& & U \sum_{\mu,i} n_{\mu,i,\uparrow}n_{\mu,i,\downarrow} + \frac{1}{2}\sum_{\mu,i \ne j} V_{i,j} n_{\mu,i}n_{\mu,j} 
\label{csterms}
\end{eqnarray}
\indent In the above $c_{\mu,i,\sigma}^{\dagger}$ creates a $\pi$-electron of spin $\sigma$ in the $p_z$ orbital of the 
$i$-th {\it C-atom} of the $\mu$-th molecular ion, $n_{\mu,i,\sigma}=c_{\mu,i,\sigma}^{\dagger}c_{\mu,i,\sigma}$, and 
$n_{\mu,i}=\sum_{\sigma}n_{\mu,i,\sigma}$. The primed sum in the site-energy dependent term of Eq.~\ref{csterms} 
is restricted to C-atoms without C-H bonds, accounting for their higher electronegativity than those with C-H bonds \cite{Huang12a}.
The nearest neighbor ($<ij>$) hopping integral of individual phenacene molecules is $t$; we take $t =$ 2.4 eV, which is the 
standard intra-molecular hopping integral for C-based systems. $U$ represents the onsite Coulomb repulsion between two electrons of 
opposite spins occupying the $2p_z$ orbital of a C-atom of a phenacene molecule; we vary $U$ from 0 $-$ 10 eV in this study.    
The parametrized expression for the $V_{i,j}$ term, which depicts the intersite repulsion between two electrons on C-atoms $i$ 
and $j$ of a phenacene molecule, is given below.
\begin{equation}  
V_{i,j} = \frac{U}{\kappa(1+0.6117R^{2}_{i,j})^{1/2}}.
\label{vij-term}
\end{equation} 
Here, $R_{i,j}$ (in \AA) is the distance between C-atoms $i$ and $j$ of a phenacene molecule. The constant $\kappa$ models the 
dielectric constant of the medium and determines the strength of $V_{i,j}$ as a function of distance \cite{Chandross97a}. 
The above parametrization with $U$ = 11.13 eV and $\kappa$ = 1 corresponds to the Ohno parametrization \cite{Ohno64a}. 

\indent
The inter-molecular Hamiltonian can include hoppping of electrons between the molecules ($H_{inter}^{1e}$), e-e repulsion between 
the charged phenacene ions ($H_{inter}^{ee}$), interactions between the metal ions and the molecules ($H_{inter}^{ion-mol}$), and 
interactions between the metal ions ($H_{inter}^{ion-ion}$). In this study we have neglected all the terms in $H_{inter}$ except
$H_{inter}^{1e}$, as they do not affect our generic qualitative theoretical model. These terms can be included later for deriving 
a quantitative theory. In what follows, 
\begin{equation}
H_{inter}^{1e} = \sum_{i \in \mu,j \in \nu,\sigma} t_{\mu,\nu,i,j} c_{\mu,i,\sigma}^{\dagger}c_{\nu,j,\sigma}.
\label{ham-inter}
\end{equation}
Here, $t_{\mu,\nu,i,j}$ are the inter-molecular hopping integrals between C-atoms $i$ and $j$ of molecules $\mu$ and $\nu$, 
respectively. Compared to the intra-molecular hopping term $t =$ 2.4 eV, the inter-molecular hoppings are tiny and generally
of the order of 0.1 $-$ 0.2 eV (cf. \onlinecite{Dutta14a}). 
\\
\indent
Solving $H$ in the complete space of $2p_{z}$ atomic orbitals is computationally impossible due to its overwhelmingly large 
dimension. We therefore work within the standard approach of molecular exciton theory and first transform $H$ from the $2p_{z}$ 
atomic orbital basis to the MO basis; this unitary basis transform is exact. We then identify the FMOs for the problem and derive 
an effective Hamiltonian in the reduced subspace of the FMOs. In the case of phenacenes, the L and L+1 orbitals accommodate the 
doped electrons upon metal-intercalation and are well separated from the remaining low and high energy MOs. In Table~\ref{tab1}
we have shown the single-particle energy gaps for phenanthrene and picene at two different values of $\epsilon$, illustrating
the separation of the FMOs from the low and high lying MOs.
 
\begin{table}[!htbp] 
\caption{Comparison of single-particle energy gaps (in units of $t$) of phenanthrene (cols 2$-$3) and picene (cols 4$-$5) 
for two different values of $\epsilon$. $H$ denote Highest Occupied MO (HOMO).}  
\begin{tabular}{| c | c | c | c | c |}
\hline
Gaps/Molecule   & $|\epsilon| =$ 0 & $|\epsilon| =$ 0.65 & $|\epsilon| =$ 0 & $|\epsilon| =$ 0.65 \\ \hline 
$\Delta_{H,L}$      &1.21   &1.20  &1.00  &0.99  \\ \hline
$\Delta_{L,L+1}$    &0.16   &0.12  &0.18  &0.13  \\ \hline
$\Delta_{L+1,L+2}$  &0.37   &0.37  &0.18  &0.19   \\ \hline
\end{tabular}
\label{tab1}
\end{table}

Due to the tiny values of inter-molecular hoppings $t_{\mu,\nu,i,j}$ relative to $\Delta_{H,L}$ and $\Delta_{L+1,L+2}$ in 
phenanthrene, occupations of the lower bonding and higher antibonding MOs are not affected by band formation involving the L and 
L+1 orbitals only. However for picene this is not the case as $\Delta_{L,L+1} \sim \Delta_{L+1,L+2}$, and participation of 
the L+2 orbital can be expected. Within the reduced subspace, due to small $\Delta_{L,L+1}$, {\it no} Coulomb (e-e) matrix 
elements between L and L+1 orbitals 
should be ignored as well as, {\it all} inter-molecular hoppings between them have to be taken into 
consideration. The importance of the inter-molecular L-L+1 hoppings will be discussed in Sec-III. The reduced Hamiltonian 
$H^{PPP}_{L,L+1}$ within the subspace of the L and L+1 orbitals can be written as,
\begin{widetext}
\begin{eqnarray}
H^{PPP}_{L,L+1} & = & \sum_{\mu,k,\sigma} \epsilon_{k} N_{\mu,k,\sigma} +  
\sum_{\mu,k,\sigma} \tilde{U}^{(d),\sigma,-\sigma}_{k,k} N_{\mu,k,\sigma} N_{\mu,k,-\sigma} +  
\sum_{\mu,k\ne k^{\prime},\sigma} \tilde{U}^{(d),\sigma,-\sigma}_{k,k{^\prime}}N_{\mu,k,\sigma} N_{\mu,k^{\prime},-\sigma} 
+ \sum_{\mu,k \ne k^{\prime} \sigma} \tilde{U}^{(d),\sigma,\sigma}_{k,k^\prime} N_{\mu,k,\sigma} N_{\mu,k^\prime,\sigma} \nonumber \\ 
              & + & 
\sum_{\mu,k_1 \neq k_2, k_3 \neq k_4,\sigma} \tilde{U}^{(o),\sigma,-\sigma}_{k_1,k_2}a_{\mu,k_1,\sigma}^{\dagger} 
a_{\mu,k_2,\sigma} a_{\mu,k_3,-\sigma}^{\dagger} a_{\mu,k_4,-\sigma}  +  
\sum_{\mu \ne \nu,k,k^\prime, \sigma} t_{\mu,\nu}^{k,k^\prime} a_{\mu,k,\sigma}^{\dagger} a_{\nu,k^\prime,\sigma}. 
\label{red}
\end{eqnarray}
\end{widetext}
Here, $k$ indices refer to L and L+1 orbitals. In the Appendix we show the derivation of $H^{PPP}_{L,L+1}$ and discuss the 
various terms occuring in reduced Hamiltonian.

\indent
We numerically study finite zig-zag 1D lattices [see Fig.~\ref{phlat}] with nearest and next-nearest neighbor hoppings between the 
molecules, using the Diagrammatic Valence Bond (DVB) technique \cite{Soos84a,SR84,SR86}. This is an exact 
diagonalization method utilizing the complete, non-orthogonal, $S_{\text{total}}$ and $S^{z}_{\text{total}}$ conserving valence 
bond basis. Within the reduced subspace of the FMOs, the real space nearest and next-nearest neighbor hoppings translate to 3 
kinds of inter-molecular inter-orbital hoppings, namely, $t_{j}^{L,L}$, $t_{j}^{L,L+1}$ and $t_{j}^{L+1,L+1}$; $j = 1,2$, denote 
nearest and next-nearest neighbors. 
The parameters of our reduced Hamiltonian [Eq.~\ref{red}] are $\Delta_{L,L+1}$, intra-molecular e-e interactions, i.e., $U$ and 
$\kappa$, and the multiple inter-molecular hopping integrals. 
While the first two are obtained from molecular calculations, we parametrize the
inter-molecular hoppings based on our previous work \cite{Dutta14a}. The parametrized values of the inter-molecular hopping 
integrals (in eV) considered in this work are as follows: $t_{1}^{L,L}$ = 0.15, $t_{1}^{L,L+1}$ = 0.05, 
$t_{1}^{L+1,L+1}$ = 0.10; $t_{2}^{L,L}$ = $t_{2}^{L+1,L+1}$ = 0.075, $t_{2}^{L,L+1}$ = 0.025. We consider $\Delta_{L,L+1}$ =
0.3 eV, 0 $\le U \le$ 10 (in eV), and $\kappa$ = 1.0, 3.0. We study 1D phenanthrene 
lattices with 10 ions (20 MOs) with charges $-$1 (10 electrons) and $-$3 (30 electrons), and 8 ions (16 MOs) with charge 
$-$2 (16 electrons). Studying 10 ions with molecular charge $-$2, i.e., 20 electrons on 20 MOs, is beyond current computational 
capacity. We compute the ratio of the average charge densities on the L and L+1 orbitals, i.e., $\frac{n_{L+1}}{n_{L}}$, and 
spin-spin correlation functions given by $\langle S^{z}_{1,k}S^{z}_{\mu,k} \rangle$, respectively; $k$ = L, L+1 and $\mu =$
molecule index. Apart from these observables, we also calculate the kinetic energies of the L and L+1 orbitals, 
\begin{widetext}
\begin{equation}
K_{p} = \Bigm| \sum_{\mu,\sigma}\biggl(t_{1}^{p,p}\langle a^{\dagger}_{\mu,p,\sigma} a_{\mu,p,\sigma} + \text{H.C.} \rangle + 
t_{2}^{p,p}\langle a^{\dagger}_{\mu,p,\sigma} a_{\mu,p,\sigma} + \text{H.C.} \rangle \biggr) \Bigm|.
\end{equation}
\end{widetext}
Here, $\mu$ is the index of the molecule and $p \in$ [L,L+1]; other symbols have already been defined.

\section{Summary of Previous Results}

\begin{figure}[tb]
\includegraphics[width=3.2in,height=1.6in]{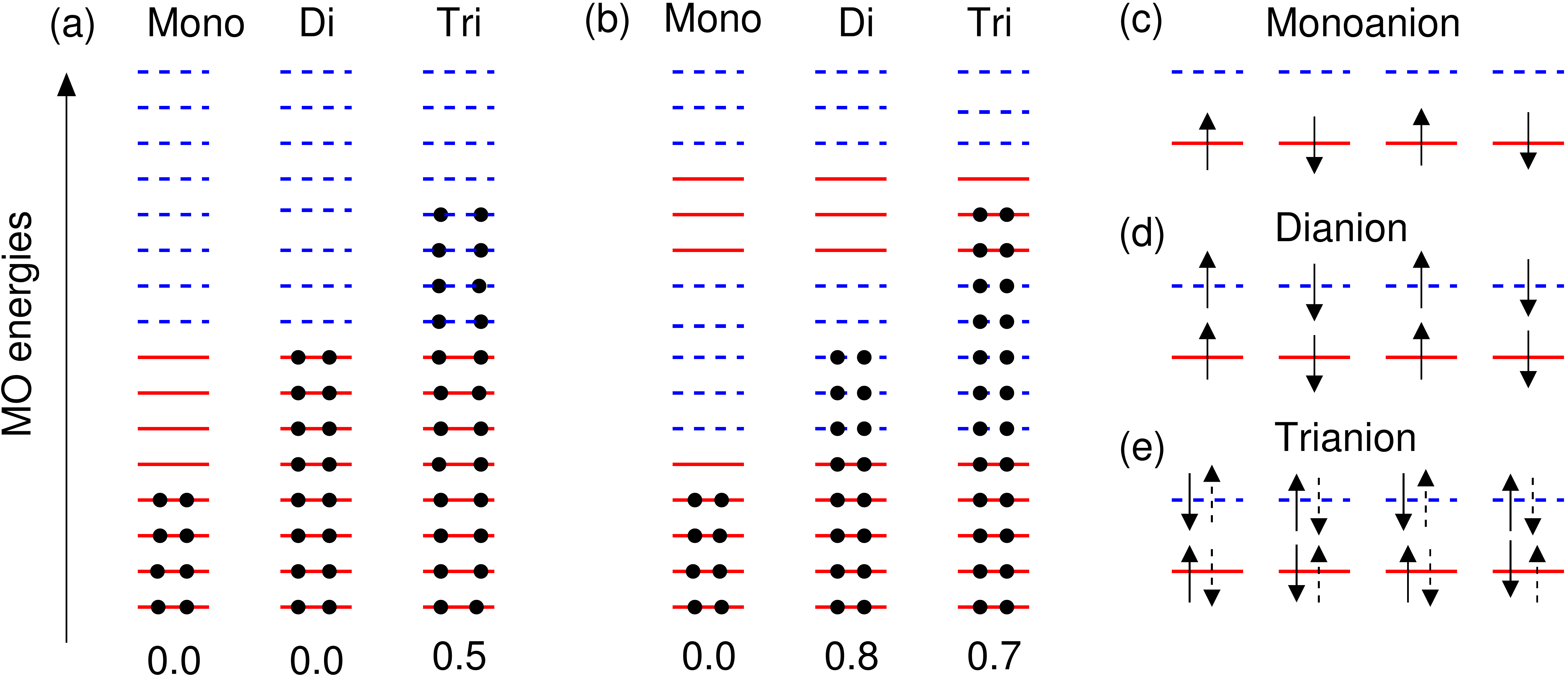}
\caption{(Color online) Tight-binding energy level diagram of a 1D lattice of $8$ phenanthrene ions, i.e., 16 MOs, without (a) and 
with (b) the $t^{L,L+1}_{j}$ terms, illustrating the role of band-width induced charge asymmetry between the mononion and trianion
crystals at $U = 0$; solid black dots indicate electrons. The numbers given below each ion in (a) and (b) indicate the ratio 
$n_{L+1}/n_{L}$; $n_{L}$ and $n_{L+1}$ are the number of electrons in the MOs with L- and L+1-character, respectively. 
Schematic representation (c)-(e) of the normal states of the three kinds of phenanthrene ionic crystals in the presence of short
range correlations; solid and broken arrows indicate $1$ and $1/2$ electron, respectively (see Sec-III for details). Solid (red) 
and broken (blue) lines indicate the L- and L+1-derived MOs, respectively.}
\label{bndstrc}
\end{figure}

Previously \cite{Dutta14a} we had derived an effective Hamiltonian, $H^{L,L+1}_{Hub}$ for metal-intercalated phenanthrene 
crystals within the localized basis of the L and L+1 orbitals of individual phenanthrene molecules. The reason for choosing
these MOs have already been discussed in the previous section. The individual phenanthene molecules were described by the Hubbard
model with short range (onsite) e-e interactions. In the $U = 0$ limit, the band structure of $H^{L,L+1}_{Hub}$ was composed of 
MOs of predominantly either L- or L+1-character. In both our past and present study, we call these as L- and L+1-derived
MOs. We found that in the {\it absence} of the $t_{j}^{L,L+1}$ terms, the band structure of $H_{Hub}^{L,L+1}$ consisted of 
L-derived MOs lying below the Fermi level and L+1-derived MOs above the Fermi level [see Fig.~\ref{bndstrc}(a)]; the Fermi level 
lies between the HOMO and the LUMO of the phenanthrene crystal with $E_{F} = 0$. When electrons were filled in this band structure, 
we found that the monoanion and the dianion crystals of phenanthrene had all L-derived MOs occupied {\it only}. As a result 
$n_{L+1}/n_{L} = 0$ for both, where $n_{L}$ and $n_{L+1}$ were the average orbital occupancies of the L- and L+1-derived MOs. 
In the trianion crystal however, electrons occupy both the L- and L+1-derived MOs, thereby making $n_{L+1}/n_{L} = 0.5$. 
In the {\it presence} of the $t_{j}^{L,L+1}$ terms however, we found that in the uncorrelated limit, both in 1D and 2D, the band 
structure of phenanthrene crystals consist of clusters of L- and L+1-derived MOs occuring alternatingly as shown in 
Fig.~\ref{bndstrc}(b). As a result, the ratio $n_{L+1}/n_{L} >$ $0$ and $0.5$ in the dianion and trianion crystals, respectively. 
The monoanion by virtue of its molecular charge, continued to have all L-derived MOs occupied only, thereby making 
$n_{L+1}/n_{L} = 0$. 
Therefore in the $U =0$ limit, presence of the $t^{L,L+1}_{j}$ terms enforced charge asymmetry between 
the monoanion and trianion crystals. {\it Note that there is no charge asymmetry between molecular valences $-1$ and $-3$ in the 
absence of the $t_{j}^{L,L+1}$ terms and both are $\frac{1}{2}$-filled systems.} The band-width effect was further enhanced by 
short range e-e interactions, 
resulting in distinct charge-orbital orderings in the three kinds of phenanthrene ionic crystals. In the dianion 
crystals the L- and L+1-derived MOs became equally populated with L-L+1 intra-molecular ferromagnetic spin ordering and, L-L 
and L+1-L+1 inter-molecular 
antiferromagnetic spin-spin couplings. The trianion crystal on the other hand behaved as a nearly 
degenerate $\frac{3}{4}$-filled two-band system with no spin order. The monoanion however, continued to be $\frac{1}{2}$-filled 
with antiferromagnetic coupling between the filled L-derived MOs. 

\section{Results and Discussion}

\begin{figure}[tb]
\includegraphics[width=3.5in]{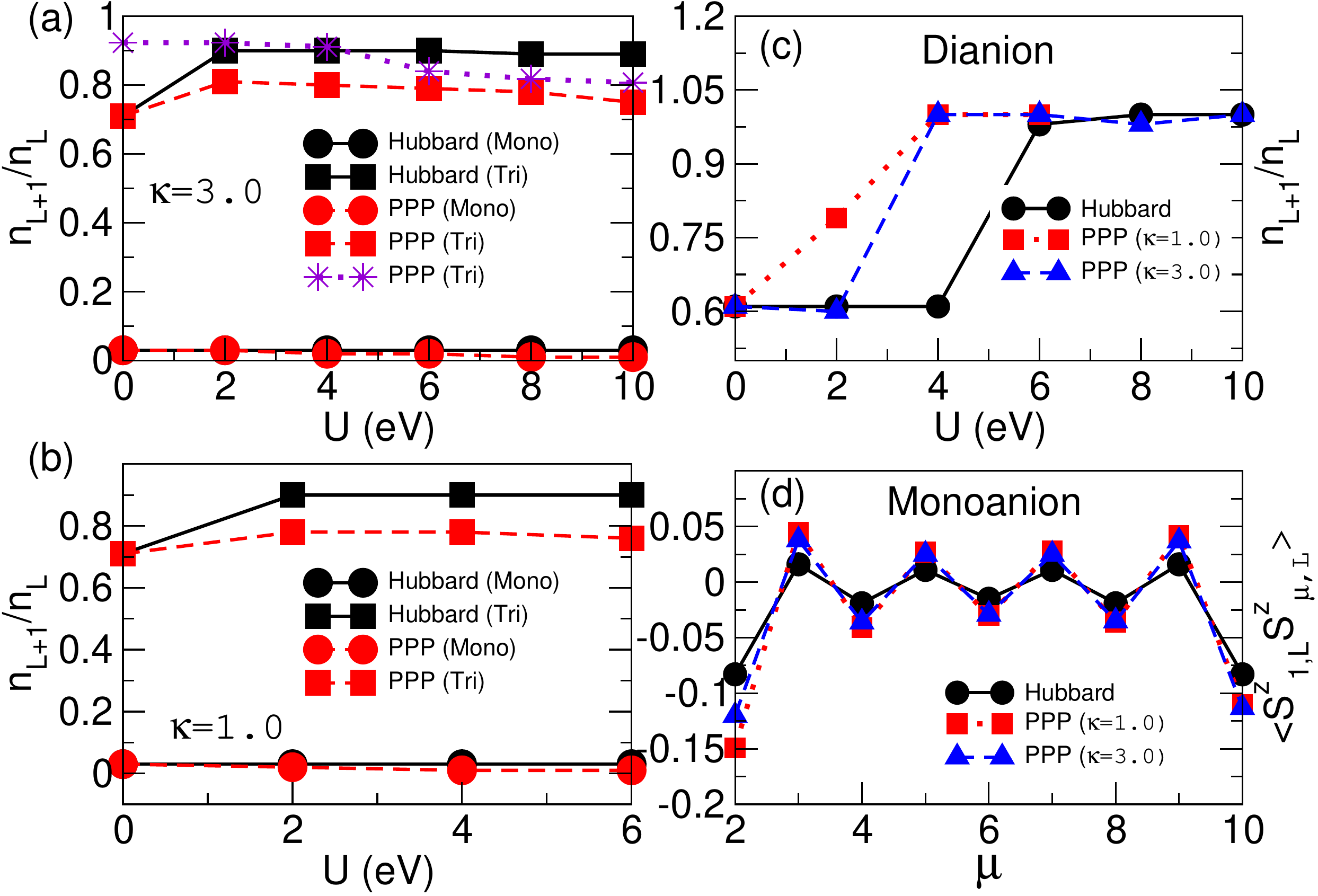}
\caption{(Color online) (a) and (b): Variation in $\frac{n_{L+1}}{n_{L}}$ with $U$ in the models $H^{Hub}_{L,L+1}$ (black solid 
lines) and $H^{PPP}_{L,L+1}$ (red dashed lines), for the monoanion (circles) and trianion (squares) crystals, respectively. Violet 
dotted line with stars in (a) is for the trianion crystal with different set of hopping parameters (see text for details). 
(c): Change in $\frac{n_{L+1}}{n_{L}}$ as a function of $U$ for the dianion crystal; solid black, dotted red and dashed blue curves 
correspond to $H^{Hub}_{L,L+1}$ model, and $H^{PPP}_{L,L+1}$ model with $\kappa = $ 1, 3, respectively. (d): Spin-spin 
correlation functions in the monoanion for the two models at $U =$ 6.0 eV; $\mu$ is the index of the molecule. L-L+1 and L+1-L+1 
spin-spin correlations being zero, are not shown. In all the plots $\Delta_{L,L+1}$ = 0.3 eV.} 
\label{Fig1Duttaphen}
\end{figure}

In Sec-III we discussed the effect of short range e-e interactions on the normal state of metal-intercalated phenanthrenes. 
Here we present our results showing the effect of long range Coulomb correlations on the same. For simplicity's sake, we will be 
referring to the diagonal e-e repulsion terms between opposite spin electrons as $\tilde{U}_{L,L}^{\uparrow,\downarrow}$,
$\tilde{U}_{L+1,L+1}^{\uparrow,\downarrow}$ and $\tilde{U}_{L,L+1}^{\uparrow,\downarrow}$. The diagonal term 
$\tilde{U}_{L,L}^{(d),\sigma,\sigma}$ representing e-e repulsion between same spin electrons will be denoted as
$\tilde{U}_{L,L+1}^{\uparrow,\uparrow}$ while the off-diagonal component $\tilde{U}_{L,L+1}^{(o),\sigma,-\sigma}$ will be 
denoted as $\tilde{U}_{L,L+1}^{2e}$. In Table~\ref{tab2} we have compared the magnitudes of the various e-e interaction 
terms present in $H^{Hub}_{L,L+1}$ and $H^{PPP}_{L,L+1}$, at two different values of $U$ and $\Delta_{L,L+1} =$ 0.3 eV. 
We find that i) magnitudes of all e-e interaction terms in $H^{PPP}_{L,L+1}$ are larger than those in $H^{Hub}_{L,L+1}$ due to 
renormalization, ii) magnitude of $\tilde{U}_{L,L+1}^{\uparrow,\downarrow}$ term in $H^{PPP}_{L,L+1}$ is either 
comparable to or larger than both $\tilde{U}_{L,L}^{\uparrow,\downarrow}$ and $\tilde{U}_{L+1,L+1}^{\uparrow,\downarrow}$
depending on the screening constant, and iii) the repulsion term $U_{L,L+1}^{\uparrow,\uparrow}$ is found to be comparable to 
$\tilde{U}_{L,L}^{\uparrow,\downarrow}$ and $\tilde{U}_{L+1,L+1}^{\uparrow,\downarrow}$ for smaller values of $\kappa$ and 
larger values of $U$. We also find that when $U > 6$ eV and $\kappa = 1$, the e-e interaction parameters in 
$H^{L,L+1}_{PPP}$ acquire unphysical values. Consequently in Fig.~\ref{Fig1Duttaphen}(b) the X-axis is truncated at $U$ = 6.0 eV. 
Based on Table~\ref{tab2} we now explain the behavior of the normal states of the three kinds of phenanthrene ionic 
crystals in the presence of long range e-e correlations.
 
\begin{table}[!htbp] 
\caption{Comparison of the magnitudes of various e-e interaction terms in $H^{Hub}_{L,L+1}$ and $H^{PPP}_{L,L+1}$ for $\kappa =$ 1, 
3, at $U$ = 4.0 eV (columns 2$-$4) and 8.0 eV (columns 5$-$7) and $\Delta_{L,L+1}$ = 0.3 eV.}
\begin{tabular}{| c | c | c | c | c | c| c| }
\hline
Param & $H^{Hub}_{L,L+1}$ & $\kappa = 1$ & $\kappa = 3$ & $H^{Hub}_{L,L+1}$ & $\kappa = 1$ & $\kappa = 3$ \\ \hline
$\tilde{U}_{L,L}^{\uparrow,\downarrow}$ &0.443 &0.789 &0.558 &0.886 &1.579 &1.117 \\ \hline
$\tilde{U}_{L+1,L+1}^{\uparrow,\downarrow}$ &0.429 &0.787 &0.548 &0.859 &1.573 &1.098 \\ \hline
$\tilde{U}_{L,L+1}^{\uparrow,\downarrow}$ &0.161 &0.935 &0.419 &0.322 &1.869 &0.838 \\ \hline
$\tilde{U}_{L,L+1}^{\uparrow,\uparrow}$ &0.000 &0.774 &0.258 &0.000 &1.547 &0.516 \\ \hline
$\tilde{U}_{L,L+1}^{2e}$ &0.161 &0.134 &0.152 &0.322 &0.267 &0.304 \\ \hline
\end{tabular}
\label{tab2}
\end{table}

\begin{figure}[tb]
\includegraphics[width=3.5in]{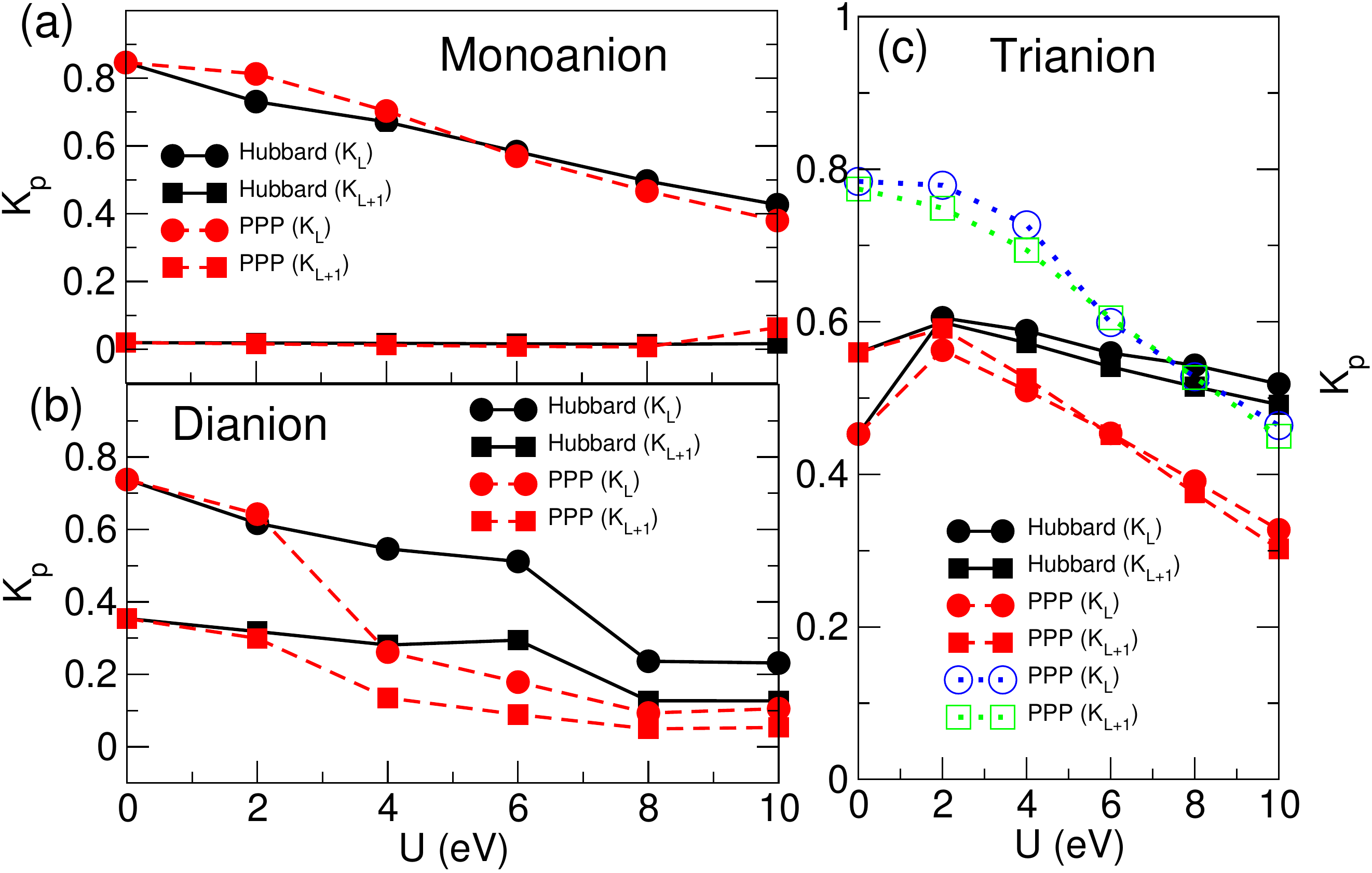}
\caption{(Color online) Change in $K_{L}$ (circles) and $K_{L+1}$ (squares) with $U$ for the monoanion (a), dianion (b) and 
trianion (c) crystals. Solid (dashed) black (red) curves pertain to $H^{Hub}_{L,L+1}$ ($H^{PPP}_{L,L+1}$) model. In all the plots 
$\Delta_{L,L+1} =$ 0.3 eV and $\kappa =$ 3.0. In (c), the dotted blue and green lines with open circles and squares, respectively, 
are for a different set of inter-molecular hoppings (see text for details).}
\label{Fig4Duttaphen}
\end{figure}

\indent
As seen from Figs.~\ref{Fig1Duttaphen}(a) and (b), the ratio $n_{L+1}/n_{L}$ in the monoanion crystal is $0$, indicating 
that the populations of the L- and L+1-derived MOs are respectively, 
$\cdots 1111 \cdots$ and $\cdots 0000 \cdots$; $1$ and $0$ refer to
singly occupied and empty sites, respectively. Thus with long range interactions also, the electrons in the monoanion crystal are 
''confined'' to the L-derived orbitals only. We therefore expect, as we had 
observed for $H_{L,L+1}^{Hub}$, that the electrons will be 
antiferromagnetically coupled to each other as this is the most stable configuration. Plot of inter-molecular L-L spin-spin 
correlations at $U =$ 6.0 eV and $\Delta_{L,L+1} =$ 0.3 eV confirms our expectation and indicate antiferromagnetic (AFM) spin 
ordering in the monoanion [Fig.~\ref{Fig1Duttaphen}(d)]. The antiferromagnetic interactions between the L-derived MOs are however 
stronger than that found with short range correlations, indicating that long range correlations stablize the $\frac{1}{2}$-filled 
AFM state more than short range interactions. This is because of increased magnitude of the $U_{L,L}^{\uparrow,\downarrow}$ term 
due to renormalization [see Table~\ref{tab2}]. Note that smaller values of the screening constant $\kappa$ sustains larger AFM 
couplings because the intra-molecular Coulomb potential $V_{i,j}$ becomes ``steeper'' with decrease in $\kappa$. As the the 
L-derived orbitals are occupied only, $K_{L}$ is non-zero as seen from Fig.~\ref{Fig4Duttaphen}(a). Due to electron hoppings, the 
site charge densities of the L-derived orbitals changes from $\cdots 1 1 1 1 \cdots$ to $\cdots 1 2 0 1 \cdots$ to 
$\cdots 1 2 1 0 \cdots$; here, $2$ refers to doubly occupied sites. The tendency to form $2s$ and $0s$ will decrease with increase 
in $U$, which dictates all the e-e repulsion terms in $H^{PPP}_{L,L+1}$. Therefore in Fig.~\ref{Fig4Duttaphen}(a) we find that 
with increase in $U$, $K_{L}$ decreases nearly monotonically. Thus, from Figs.~\ref{Fig1Duttaphen}(a), (b) and (d), and 
Fig.~\ref{Fig4Duttaphen}(a) we find that the behavior of the monoanion crystal is the same with both short and long range e-e 
interactions. We therefore surmise that long range correlations do not affect the normal state of the monoanion crystal and it 
continues to behave as a $\frac{1}{2}$-filled Mott-Hubbard semiconductor.     

\begin{figure}[tb]
\includegraphics[width=3.5in]{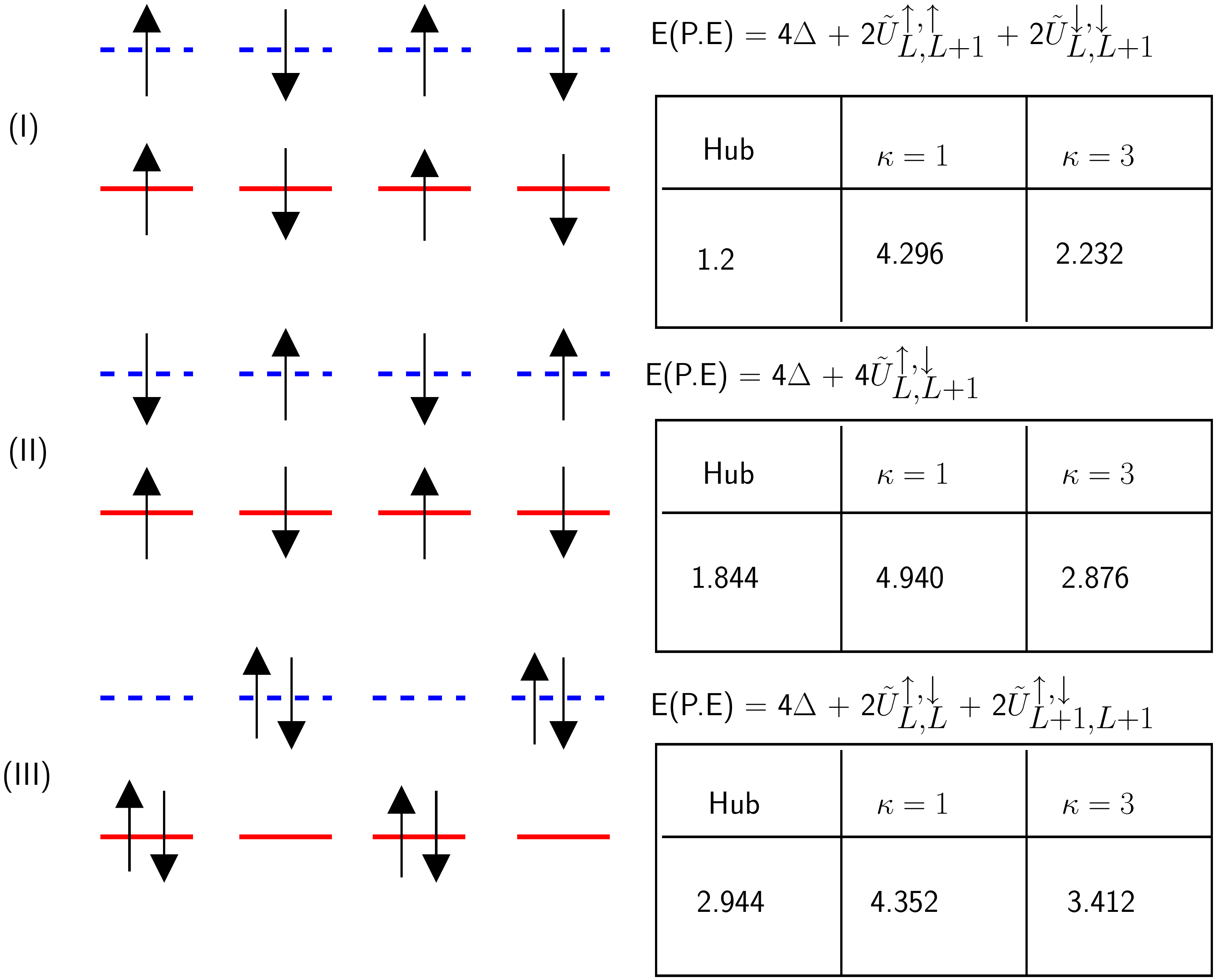}
\caption{(Color online) Schematic depiction of three configurations of the dianion crystal with equally populated L- (solid red
line) and L+1-derived (dashed blue line) orbitals. The analytical expression of the potential energy, $E(P.E)$, of each 
configuration is indicated along with its numerical value in the table underneath, for $H_{L,L+1}^{Hub}$ and $H_{L,L+1}^{PPP}$ 
($\kappa =$ 1, 3). Arrows denote electrons, $U =$ 4.0 eV, and $\Delta = \Delta_{_L,L+1} =$ 0.3 eV. The numerical values of $E(P.E)$ 
are obtained using the analytical expression and Table~\ref{tab2}}
\label{di}
\end{figure}

\begin{figure}[tb]
\includegraphics[width=3.5in]{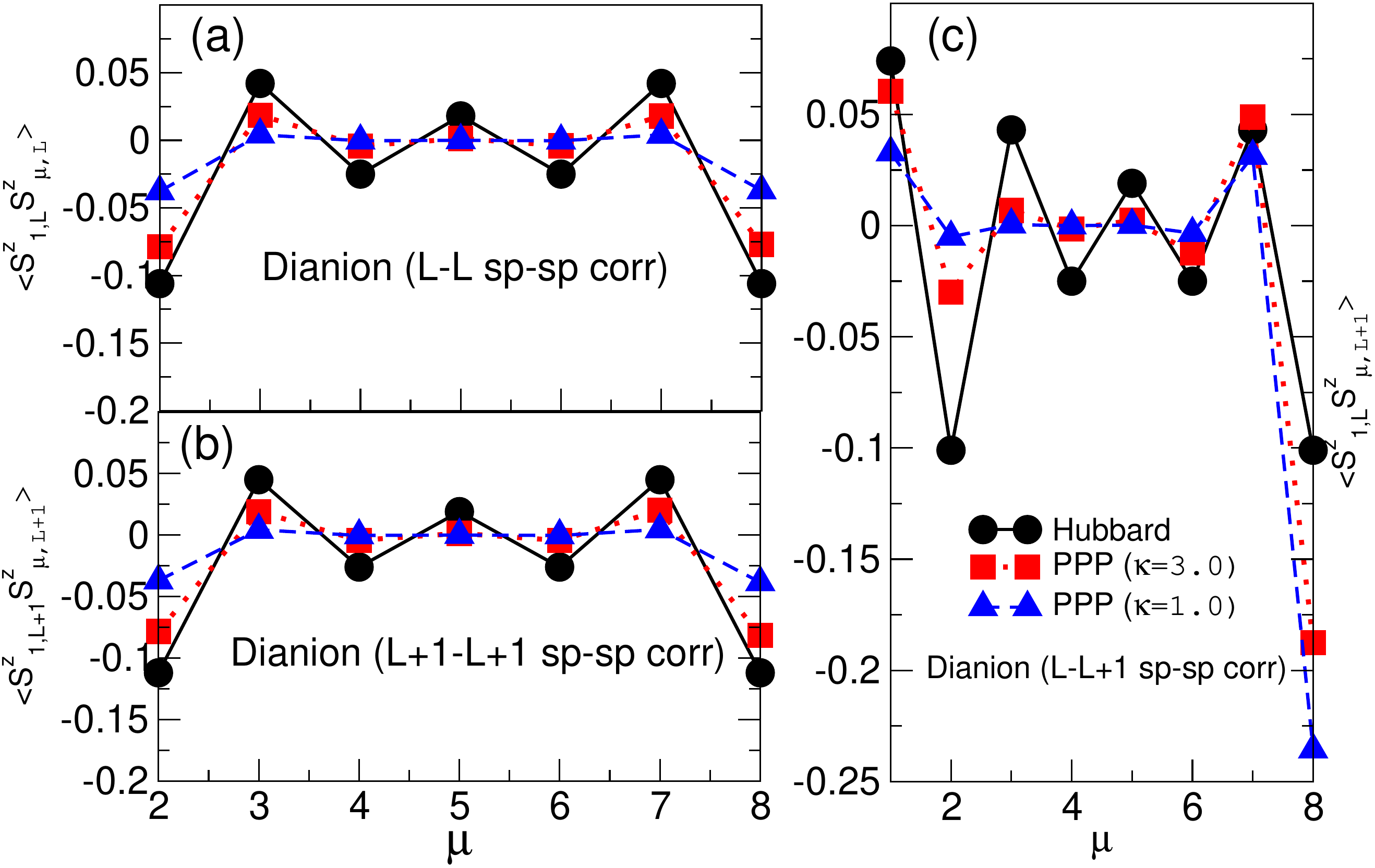}
\caption{(Color online) Spin-spin correlations for the dianion crystal in the $H_{L,L+1}^{Hub}$ model (solid black line with 
circles) and the $H_{L,L+1}^{PPP}$ model with $\kappa =$ 1 (dotted red line with squares) and $\kappa =$ 3 
(dashed blue line with triangles), respectively. (a), (b) and (c) show the plots of $\langle S^{z}_{1,L} S^{z}_{\mu,L} \rangle$, 
$\langle S^{z}_{1,L+1} S^{z}_{\mu,L+1} \rangle$, and $\langle S^{z}_{1,L} S^{z}_{\mu,L+1} \rangle$, respectively, for 
$\Delta_{L,L+1} =$ 0.3 eV and $U =$ 6.0 eV; $\mu$ is the index of the molecule.}
\label{Fig2Duttaphen}
\end{figure}

\indent
We have already seen in Sec-III that the band structure of the dianion crystal is such that it allows electron exchange between 
the L- and L+1-derived MOs at $U = 0$. Population exchange between these MOs is further enhanced by short range e-e 
interactions \cite{Dutta14a} resulting in $n_{L+1}/n_{L} \rightarrow 1$. From Fig.~\ref{Fig1Duttaphen}(c) it is observed that with 
long range correlations also, as $U$ is increased, $n_{L+1}/n_{L} \rightarrow 1$. However, compared to short range interactions, 
long range correlations bring about population equalization ``faster'', i.e., at relatively smaller $U$ values. Furthermore, the 
atomic $U$ at which $n_{L+1}$ becomes equal to $n_{L}$ is dictated by the screening constant, i.e., the ``steepness'' of the 
intra-molecular potential $V_{i,j}$. $n_{L+1}/n_{L} \rightarrow 1$ in the dianion crystals in the presence of long range 
correlations is very intriguing because the repulsion term $\tilde{U}_{L,L+1}^{\uparrow,\uparrow}$ is supposed to hinder 
population equalization. To understand this behavior of dianion crytals, we compare the potential energies, $E(P.E)$, of three 
configurations with equally populated L- and L+1-derived MOs in Fig.~\ref{di}. We choose these because configurations with equally 
populated L- and L+1-derived orbitals have maximum kinetic stability and are thus prone to be favored over other configurations. 
From Fig.~\ref{di} it is evinced that configurations like (I) have the minimum potential energy and are hence dominant in the ground 
state of the dianion crystal. This explains the behavior of the spin-spin correlation functions shown in Fig.~\ref{Fig2Duttaphen}: 
While the inter-molecular L-L and L+1-L+1 spin-spin correlations are antiferromagnetic in nature, the intra-molecular L-L+1 
spin-spin couplings are ferromagnetic. Depending on the magnitude of $U$ the dianion crystal behaves as a two-band system with 
either 0 $< n_{L+1}/n_{L} <$ 1 or $n_{L+1}/n_{L} =$ 1. These two ground states being different from each other, are expected to have 
different kinetic energies. This is indeed the case as seen from Fig.~\ref{Fig4Duttaphen}(b). If we compare 
Figs.~\ref{Fig1Duttaphen}(c) and \ref{Fig4Duttaphen}(b) it is found that the $U$ value at which $n_{L+1}/n_{L}$ increases suddenly 
coincides with the $U$ value at which $K_{L}$ and $K_{L+1}$ show sudden decrease. Once population equalization has occurred, 
$K_{L} \simeq K_{L+1}$ and both of them decrease monotonically with $U$ as the L- and L+1-derived orbitals become effectively 
$\frac{1}{2}$-filled. Due to finite size effect the values of $K_{L}$ and $K_{L+1}$ are not of same magnitudes although in the 
thermodynamic limit they are expected to be same. 

\begin{figure}[tb]
\includegraphics[width=3.5in]{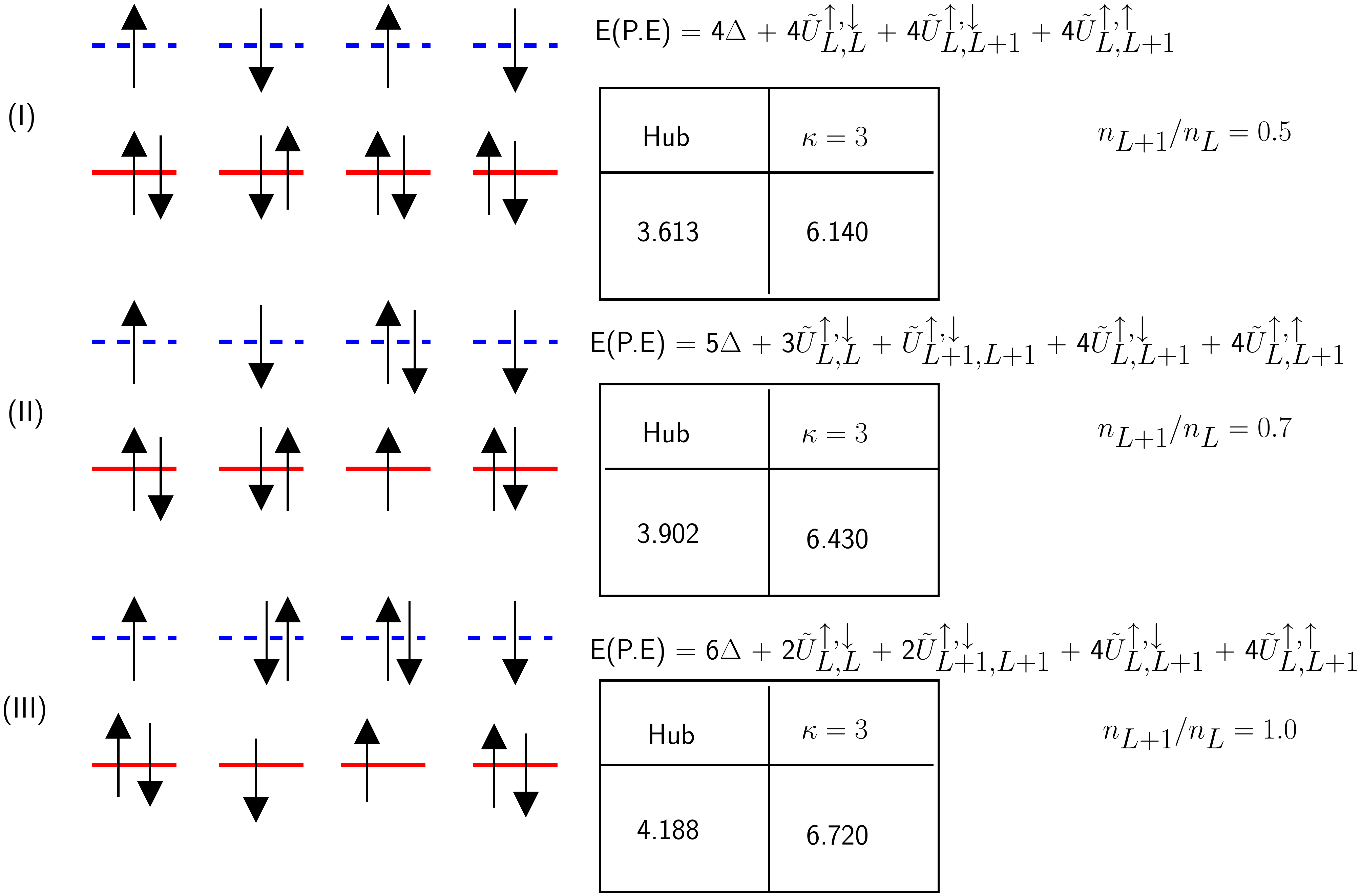}
\caption{(Color online) Schematic representation of three configurations of the trianion crystal along with their corresponding 
potential energies, $E(P.E)$. The table for each configuration depict the numerical values of $E(P.E)$ for $H^{Hub}_{L,L+1}$ and 
$H^{PPP}_{L,L+1}$ ($\kappa =$ 1, 3), respectively, at $U = 4$ eV; these values have been obtained using the analytical expressions 
of $E(P.E)$ and Table~\ref{tab2}. Solid red and dashed blue lines depict the L- and L+1-derived MOs, respectively; $\uparrow$ 
($\downarrow$) denote spin-up (spin-down) electrons. The ratio $n_{L+1}/n_{L}$ for each configuration is also shown. 
(I) is the Mott-Hubbard configuration, (II) is the $U = 0$ configuration, and (III) is the $\frac{3}{4}$-filled two-band system.} 
\label{tri}
\end{figure}

\begin{figure}[tb]
\includegraphics[width=3.5in]{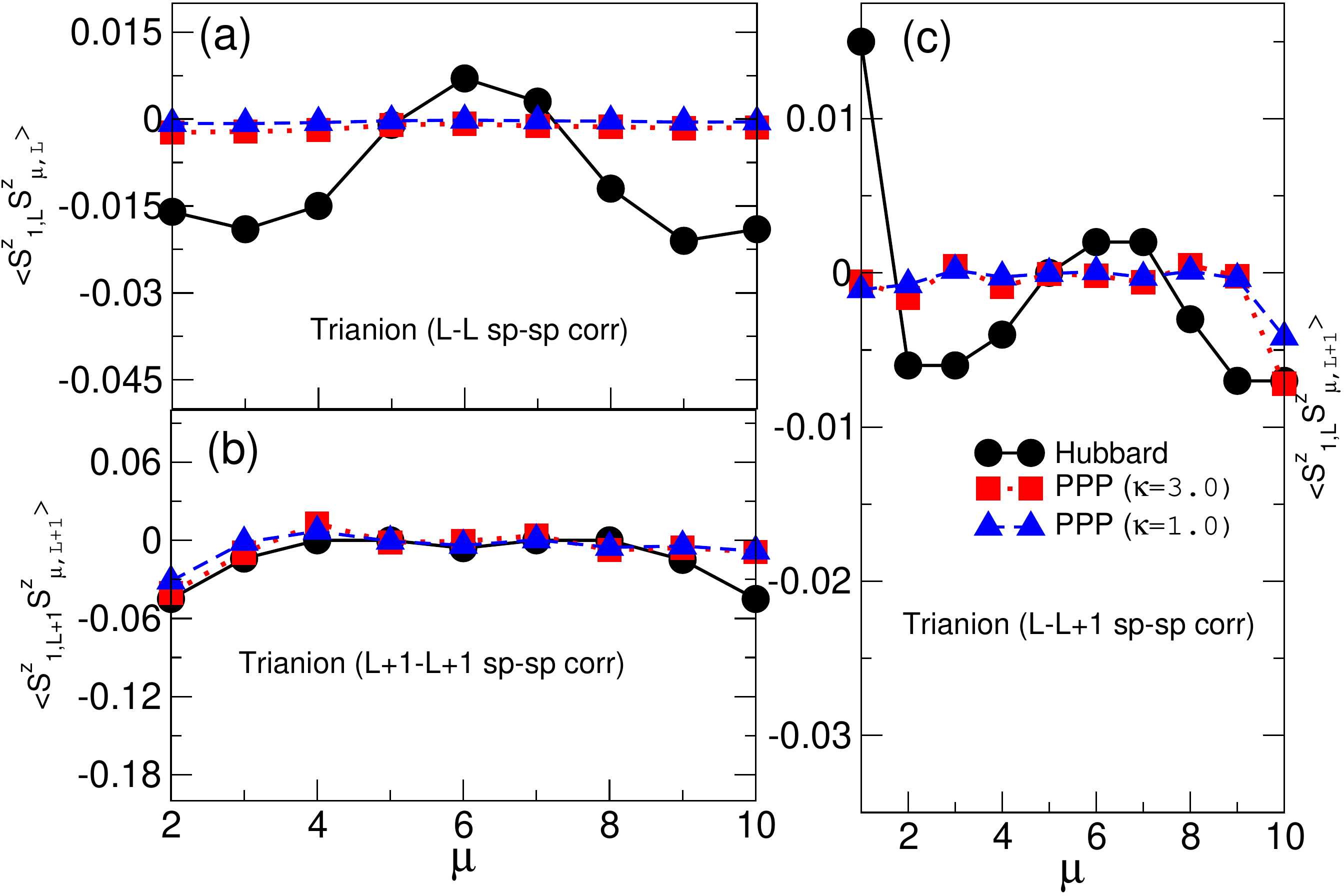}
\caption{(Color online) Change in $\langle S^{z}_{1,L} S^{z}_{\mu,L} \rangle$ (a), 
$\langle S^{z}_{1,L+1} S^{z}_{\mu,L+1} \rangle$ (b) and $\langle S^{z}_{1,L} S^{z}_{\mu,L+1} \rangle$ (c), with $U$ of the 
trianion crystal at $\Delta_{L,L+1}$ = 0.3 eV and $U =$ 6.0 eV; $\mu$ is the molecule index. 
Solid black lines with circles indicate the $H^{Hub}_{L,L+1}$ model while dotted (dashed) red (blue) curves with 
squares (triangles) refer to $H^{PPP}_{L,L+1}$ with $\kappa =$ 1 ($\kappa =$ 3), respectively.}
\label{Fig3Duttaphen}
\end{figure}

\indent
By virtue of the band-width effect, crystals of phenanthrene trianions have $n_{L+1}/n_{L} >$ 0.5 at $U =$ 0. 
From  Figs.~\ref{Fig1Duttaphen}(a) and (b) it is observed that short range e-e interactions further enhance this charge 
transfer between the L- and L+1-derived orbitals, thereby making $n_{L+1}/n_{L} \rightarrow$ 1.0.  
Long range correlations on the other hand suppress this tendency and coerce $n_{L+1}/n_{L}$ to approach its $U =0$ value with 
increase in $U$, as evinced from the same figures. In other words, the tendency of trianion crystals to form a nearly-degenerate 
$\frac{3}{4}$-filled system is hindered by long range correlations. In order to understand this, we will first briefly discuss 
how short range e-e interactions drive trianion crystals toward population equalization. Consider the Mott-Hubbard configuration 
[configuration (I) in Fig.~\ref{tri}]. Although it has the minimum potential energy compared to all other configurations, its 
kinetic stability is also minimum. This is because of $K_{L}$ being zero. Any other configuration with $n_{L+1}/n_{L} >$ 0.5 has 
more kinetic stability due to contribution from both $K_{L}$ and $K_{L+1}$. This gain in kinetic energy overrides the increase in 
potential energy due to electron exchange between the L- and L+1-derived orbitals. Now, consider configurations (II) and (III) 
of Fig.~\ref{tri}; while (II) has $n_{L} > n_{L+1}$, (III) has $n_{L} = n_{L+1}$. Although the potential energies of these 
configurations are comparable, any {\it double occupancy conserving} electron hops for configuration (II) will lead to an increase 
in potential energy. No potential energy increase will however occur for configuration (III) by virtue of these {\it double
occupancy conserving} hops; we define {\it double occupancy conserving} hoppings as electron hops within the L-derived
(L+1-derived) orbitals such that the number of doubly occupied L-derived (L+1-derived) orbitals remain unchanged \cite{Dutta14a}. 
In other words, configurations with $n_{L+1}/n_{L} \simeq$ 1 will have higher kinetic stability and therefore dominate in the 
ground state wavefunction at $U >$ 0.    

\indent
With long range correlations however, configurations like (II) dominate the ground state wavefunction and therefore 
$n_{L+1}/n_{L}$(at $U > 0$) $\rightarrow n_{L+1}/n_{L}$(at $U = 0$). 
This implies that the {\it double occupancy conserving} electron hops fail 
to provide configurations like (III) the kinetic stability needed to overcome the initial potential energy gain. If we look at the 
magnitudes of $E(P.E)$ in Fig.~\ref{tri}, we find that magnitudes of the potential energies of the configurations in the 
presence of long range
correlations are almost double compared to their magnitudes with short range correlations. This indicates that the behavior of
the trianion crystals with long range e-e interactions will be dominated by $E(P.E)$ rather than kinetic 
stability. Simply stated, this implies that with increase in $U$ the ground state wave function of the trianion crystal will be 
dominated by configurations which are kinetically more stable than the Mott-Hubbard configuration but with the minimum number of 
electrons promoted from the L- to the L+1-derived orbitals. For example, in the lattice of 20 MOs with 30 electrons, for 
$0 < U \le 10$, only 2 electrons are transferred from the L- to the L+1-derived MOs, thereby making $n_{L+1}/n_{L} \sim 0.7$ 
[see Figs.~\ref{Fig1Duttaphen}(a) and (b)]. We believe that the behavior of the trianion crystal in the presence of long range 
correlations is due to smaller band-width compared to the e-e interactions. In case of short range correlations the band-width
was comparable to the e-e interactions and thus allowed cooperation between the two. There are two possible ways to increase the
band-width: (a) Increase the magnitudes (within realistic limits) 
of the inter-molecular inter-orbital hopping terms, i.e., $t_{j}^{\mu,\nu}$, and, (b) 
increase the lattice sizes chosen. However, (b) is not possible in the present context as the lattices used are the largest possible 
within our exact diagonalization study. Therefore if the behavior of the trianion crystal does not change upon increasing the 
magnitudes of the $t_{j}^{\mu,\nu}$ terms, then small system size is the reason for the trianion crystal for behaving
differently in the presence of long range correlations.

\indent
We computed $n_{L+1}/n_{L}$ and $K_{L}$ and $K_{L+1}$ for the trianion crystal with larger hopping parameters, i.e., larger 
band-width, at $\Delta_{L,L+1} =$ 0.3 eV 
and $\kappa =$ 3.0; $t_{1}^{L,L} =$ 0.175, $t_{1}^{L,L+1} =$ 0.075, $t_{1}^{L+1,L+1} =$ 0.125, and, 
$t_{2}^{L,L} = t_{2}^{L+1,L+1} =$ 0.1, $t_{2}^{L,L+1} =$ 0.05; all hopping terms are realistic and expressed in eV. The plots of 
$n_{L+1}/n_{L}$ and the kinetic energies versus $U$ with the new $t^{k,k^\prime}_{j}$ terms are shown in 
Figs.~\ref{Fig1Duttaphen}(a) and \ref{Fig4Duttaphen}(c), 
respectively. We find that the behavior of the trianion crystal at large $U$ remains unchanged even with increased hopping 
parameters. We therefore conclude that long range interactions demand longer system sizes compared to those studied in the present 
work. The L-L, L+1-L+1 and L-L+1 spin-spin correlations [Figs.~\ref{Fig3Duttaphen}(a)-(c)] however show that there is no spin 
ordering in the trianion as was found with short range correlations. 
This again indicates that the ground state of the trianion is not of the 
Mott-Hubbard type with $n_{L} =$ 2 and $n_{L+1} =$ 1, but with $n_{L}$ and $n_{L+1}$ slightly less and more than 2 and 1, 
respectively. 

\section{Conclusions and Discussions}

In this work we have addressed the effect of long range Coulomb correlations on the normal state of metal-intercalated 
phenacene crystals, taking phenanthrene as example. The individual phenacene molecules have been modeled by the Pariser-Parr-Pople 
Hamiltonian. We derive an effective correlated model Hamiltonian $H^{PPP}_{L,L+1}$ to describe the phenacene ions 
in the solid state within a localized basis of the L and L+1 FMOs. We find that long range correlations not only renormalize the 
Coulomb terms occuring in $H^{Hub}_{L,L+1}$ but also introduce an additional intra-molecular repulsion term between same spin 
electrons; $H^{Hub}_{L,L+1}$ is the effective Hamiltonian obtained by describing the individual phenacene molecules by the 
Hubbard model with short range e-e interactions. It is found that the normal states of the monoanion and dianion molecular solids 
remain unaffected by the presence of long range interactions. The monoanion crystal continues to behave as a single-band 
$\frac{1}{2}$-filled Mott-Hubbard semiconductor with L-L antiferromagnetic spin order. The dianion crystal undergo population 
equalization with increase in the strength of e-e repulsion, i.e., $U$. Spin-spin correlations show that the 
inter-molecular L-L and L+1-L+1 spin couplings are antiferromagnetic while the intra-molecular L-L+1 spin couplings are 
ferromagnetic. The trianion's behavior with short and long range interactions is however different. With short range
correlations the trianion is a nearly $\frac{3}{4}$-filled two-band system without any spin order. However, in the presence of long 
range interactions, with increase in $U$, the tendency towards population equalization between the L- and L+1-derived MOs is 
suppressed and the ratio $n_{L+1}/n_{L}$ reaches its $U = 0$ magnitude. This occurs because the system sizes chosen in
this study are not large enough to provide the kinetic stability necessary for $n_{L+1}/n_{L} \simeq 1$. 
Thus the trianion crystal behaves as a two-band system with $n_{L}$ slightly less than 2 and $n_{L+1}$ slightly greater than 1,
and without any spin order. We believe that larger system sizes will allow population equalization in the trianion crystal 
even with long range correlations. We thus conclude that the behavior of the normal state of metal-intercalated phenanthrene
crystals are not affected by long range Coulomb correlations.

\indent
Occurance of SC under pressure and at a fixed carrier concentration is a signature of C-based superconductors. While SC occurs at 
carrier density $\rho =$ 0.5 in the CTS, both fullerides and metal-intercalated phenacenes show SC at the molecular stoichimetry 
of $-3$. The CTS can be either electron or hole doped with one electron or hole per dimer resulting in $\rho =$ 0.5. Examples of 
hole and electron doped CTS are $\kappa-$(BEDT)$_{2}$X, where X = Cl, Br, I, (NCS)$_{2}$, Cu$_{2}$(CN)$_{3}$, and, 
Z[Pd(dmit)$_{2}$]$_{2}$ salts, Z = P, As, Sb, respectively; here, BEDT = bisethylenedithio-tetrathiafulvalene and 
dmit = 1,3-dithiol-2-thione-4,5-dithiolate. In metal-intercalated phenacenes, two MOs per molecule accommodate $3$ electrons and 
thus $\rho =$ $\frac{3}{2}$. Similar scenario is expected in the case of doped fullerides also: $3$ electrons on two non-degenerate 
MOs separated by a gap equal to the Jahn-Teller distortion energy, thereby giving $\rho = \frac{3}{2}$. However as discussed in 
Sec-I, all available theoretical works on the fullerides are based on the idea that SC occurs from an effectively 
$\frac{1}{2}$-filled ($\rho = 1$) Mott-Hubbard AFM; this theoretical approach however suffers from some drawbacks, for example, 
it is unable to explain the absence of SC in singly-doped fullerides. Thus, it is important to revisit the mechanism of SC in doped 
fullerides. Also, instead of focussing on each of these three classes of C-based superconducting materials separately, stress should 
be given on understanding whether these materials indeed have a common normal state and thus, a common mechanism of SC. 
\\

{\bf Acknowledgement}
This work was supported by the U.S. Department of Energy, Office of Science, Basic Energy Sciences, under Award No.
DE-FG02-06ER46315. This research used resources of the NERSC, which is supported by the Office of Science of the U.S. 
Department of Energy under Contract No. DE-AC02-05CH11231.
\\

\appendix
\section{Derivation of the reduced Hamiltonian}
\indent The complete Hamiltonian of metal-intercalated phenacenes in the solid state, within the $2p_{z}$ atomic orbital basis, can 
be expressed as, $H$ = $H_{intra}$ + $H_{inter}$; $H_{intra}$ [Eq.~\ref{csterms}] 
can be broken into $H_{intra}^{1e}$ and $H_{intra}^{ee}$ which are
the one- and many-electron components, respectively. In order to transform $H$ from the atomic orbital basis to the 
MO basis, we first solve $H_{intra}^{1e}$ exactly,
\begin{equation}
H_{intra}^{1e}=\sum_{\mu,k,\sigma}E_{\mu,k}a_{\mu,k,\sigma}^\dagger a_{\mu,k,\sigma}, 
\label{Huckel}
\end{equation}
where $a_{\mu,k,\sigma}^{\dagger}=\sum_{i}\psi_{\mu,k,i}c_{\mu,i,\sigma}^{\dagger}$ represents the $k$th MO of the $\mu$-th
molecule. We then express $H_{intra}^{ee}$ and $H_{inter}^{1e}$ within these {\it localized} MOs
$a_{\mu,k,\sigma}^\dagger$ \cite{Dutta14a,Chandross99a}.
\begin{widetext}
\begin{equation}
H_{intra}^{ee}(\text{onsite})= 
U \biggl[ \sum_{\mu,k,k^\prime,i} |\chi_{\mu,i,k}|^2|\chi_{\mu,i,k^{\prime}}|^2 N_{\mu,k,\uparrow} N_{\mu,k^\prime,\downarrow} + 
\sum_{\mu,k_1 \neq k_2, k_3 \neq k_4,i}(\prod_{l=1}^{4}\chi_{\mu,i,k_l})a_{\mu,k_1,\uparrow}^{\dagger} a_{\mu,k_2,\uparrow} 
a_{\mu,k_3,\downarrow}^{\dagger} a_{\mu,k_4,\downarrow} \biggr] 
\label{reduced1}
\end{equation}
\begin{eqnarray}
H_{intra}^{ee}(\text{long range}) = 
\frac{1}{2} \biggl[\sum_{\mu,k,\sigma,i\ne j} V_{i,j}|\chi_{\mu,i,k}|^{2} |\chi_{\mu,j,k}|^{2} N_{\mu,k,\sigma} &+& 
\sum_{\mu,k,k^{\prime},\sigma,i\ne j} V_{i,j} |\chi_{\mu,i,k}|^2|\chi_{\mu,j,k^{\prime}}|^2 N_{\mu,k,\sigma} 
N_{\mu,k^{\prime},-\sigma} \nonumber \\ 
+ \sum_{\mu,k \ne k^{\prime},\sigma,i\ne j} 
V_{i,j} |\chi_{\mu,i,k}|^2|\chi_{\mu,j,k^{\prime}}|^2 N_{\mu,k,\sigma} N_{\mu,k^{\prime},\sigma} 
+ \sum_{\mu,k_1 \neq k_2, k_3 \neq k_4;i\ne j} & V_{i,j} & \biggl\{\prod_{l=1}^{2}\chi_{\mu,i,k_l}\chi_{\mu,j,k_l}\biggr \} 
a_{\mu,k_1,\uparrow}^{\dagger} a_{\mu,k_2,\uparrow} a_{\mu,k_3,\downarrow}^{\dagger} a_{\mu,k_4,\downarrow} \nonumber \\
+ \sum_{\mu,k\ne k^{\prime},i\ne j,\sigma} 
\biggl\{|\chi_{\mu,i,k}|^{2}\chi_{\mu,j,k}\chi_{\mu,j,k^{\prime}}+|\chi_{\mu,j,k}|^{2}\chi_{\mu,i,k}\chi_{\mu,i,k^{\prime}} \biggr\} 
& N_{\mu,k,\sigma} & (a^{\dagger}_{\mu,k,\sigma}a_{\mu,k^{\prime},\sigma} + \text{H.C.}) \biggr] 
\label{reduced2}
\end{eqnarray}
\begin{equation}
H_{inter}^{1e}  = \sum_{k_1,k_2,\sigma;i \in \mu,j \in \nu,\mu \ne \nu} \chi_{\mu,i,k_1} \chi_{\nu,j,k_2} t_{\mu,\nu,i,j} a_{\mu,k_1,\sigma}^{\dagger} a_{\nu,k_2,\sigma} 
\label{reduced3}
\end{equation}
\end{widetext}
\indent The matrices $\chi$ and $\psi$ are inverses of each other and $N_{\mu,k,\sigma}=a_{\mu,k,\sigma}^{\dagger}a_{\mu,k,\sigma}$.
The above transformation from the complete basis of atomic orbitals to the complete MO basis is {\it exact}. The terms
$H_{intra}^{ee}$(onsite) and $H_{intra}^{ee}$(long range) correspond to the intra-molecular inter-atomic e-e repulsions due to 
the short and long range components of the PPP Hamiltonian (Eq.~\ref{csterms}). 
The second and fourth terms of Eq.~\ref{reduced2} are same as the first and second terms in Eq.~\ref{reduced1}, respectively, 
except with different coefficients. Thus, the long range component of the PPP Hamiltonian renormalize the onsite ($U$-dependent) 
terms in Eq.~\ref{reduced1}. The third term in Eq.~\ref{reduced2} denotes the e-e repulsion between {\it parallel} spins on 
different MOs, {\it within the same molecule}; the last term in Eq.~\ref{reduced2} depict density-dependent intra-molecular 
hoppings between different MOs. Eqs.~\ref{reduced1} $-$~\ref{reduced3} represent $H$ in the complete MO basis. 
The reduced Hamiltonian $H_{L,L+1}$ can be obtained by restricting the sums over the {\it k's} in 
Eqs.~\ref{reduced1} $-$~\ref{reduced3} to the L and L+1 orbitals.
\begin{widetext}
\begin{eqnarray}
H^{PPP}_{L,L+1} & = & \sum_{\mu,k,\sigma} \epsilon_{k} N_{\mu,k,\sigma} +  
\sum_{\mu,k,\sigma} \tilde{U}^{(d),\sigma,-\sigma}_{k,k} N_{\mu,k,\sigma} N_{\mu,k,-\sigma} +  
\sum_{\mu,k\ne k^{\prime},\sigma} \tilde{U}^{(d),\sigma,-\sigma}_{k,k{^\prime}}N_{\mu,k,\sigma} N_{\mu,k^{\prime},-\sigma} 
+ \sum_{\mu,k \ne k^{\prime} \sigma} \tilde{U}^{(d),\sigma,\sigma}_{k,k^\prime} N_{\mu,k,\sigma} N_{\mu,k^\prime,\sigma} \nonumber \\ 
              & + & 
\sum_{\mu,k_1 \neq k_2, k_3 \neq k_4,\sigma} \tilde{U}^{(o),\sigma,-\sigma}_{k_1,k_2}a_{\mu,k_1,\sigma}^{\dagger} 
a_{\mu,k_2,\sigma} a_{\mu,k_3,-\sigma}^{\dagger} a_{\mu,k_4,-\sigma}  +  
\sum_{\mu \ne \nu,k,k^\prime, \sigma} t_{\mu,\nu}^{k,k^\prime} a_{\mu,k,\sigma}^{\dagger} a_{\nu,k^\prime,\sigma} 
\label{H_LL+1}
\end{eqnarray}
\end{widetext}
In Eq.~\ref{H_LL+1} all $k$ indices denote L and L+1 MOs and $t_{\mu,\nu}^{L,L+1}$ = 
$\sum_{i,j}\chi_{\mu,i,k_{1}}\chi_{\nu,j,k_{2}}t_{\mu,\nu,i,j}$. The orbital (site) energies $\epsilon_{k}$ and the various  
spin-dependent Coulomb terms are defined as given below.
\begin{widetext}
\begin{eqnarray*}
\epsilon_{L} & = & E_{L} + \frac{1}{2}\sum_{i\ne j} V_{i,j}|\chi_{\mu,i,L}|^{2} |\chi_{\mu,j,L}|^{2}; 
\epsilon_{L+1} = E_{L+1} + \frac{1}{2}\sum_{i\ne j} V_{i,j}|\chi_{\mu,i,L+1}|^{2} |\chi_{\mu,j,L+1}|^{2} \\
\tilde{U}_{L,L}^{(d),\sigma,-\sigma} & = & \sum_{i}|\chi_{\mu,i,L}|^2 \biggl(U|\chi_{\mu,i,L}|^2 + \frac{1}{2}\sum_{j\ne i} V_{i,j}|\chi_{\mu,j,L}|^2 \biggr); \tilde{U}_{L+1,L+1}^{(d),\sigma,-\sigma} =  \sum_{i}|\chi_{\mu,i,L+1}|^2 \biggl(U|\chi_{\mu,i,L+1}|^2 + \frac{1}{2}\sum_{j\ne i} V_{i,j}|\chi_{\mu,j,L+1}|^2 \biggr) \\
\tilde{U}_{L,L+1}^{(d),\sigma,-\sigma} & = & \sum_{i}|\chi_{\mu,i,L}|^2 \biggl(U|\chi_{\mu,i,L+1}|^2 + \frac{1}{2}\sum_{j\ne i} V_{i,j}|\chi_{\mu,j,L+1}|^2 \biggr); \tilde{U}_{L,L+1}^{(d),\sigma,\sigma} = \frac{1}{2}\sum_{i,j\ne i}V_{i,j}|\chi_{\mu,i,L}|^2 |\chi_{\mu,j,L+1}|^2 \\
\tilde{U}_{L,L+1}^{(o),\sigma,-\sigma} & = & U \sum_{i} \chi^{2}_{\mu,i,L}\chi_{\mu,i,L+1}^{2} + \frac{1}{2}\sum_{i,j;j\ne i} V_{i,j}\chi^{2}_{\mu,i,L}\chi_{\mu,j,L+1}^{2} 
\label{exps} 
\end{eqnarray*}
\end{widetext}

\indent The diagonal Coloumb terms, $U_{k,k}^{(d),\sigma,-\sigma}$ $k \in$ [L,L+1], denote the intra-molecular Coulomb 
repulsions between two electrons of opposite spins occupying orbital $k$. When both the L and L+1 orbitals on the {\it same}
molecule are singly occupied by electrons with {\it opposite} ({\it same}) spins, the e-e repulsion is denoted by    
$U_{L,L+1}^{(d),\sigma,-\sigma}$ ($U_{L,L+1}^{(d),\sigma,\sigma}$). Apart from the four diagonal Coulomb terms, two 
off-diagonal terms of equal magnitude and denoted by $U_{L,L}^{(o),\sigma,-\sigma}$ are also present in Eq.~\ref{H_LL+1}.
One of them is refered to as the two-electron (2e) hopping term and is responsible for promoting an electron pair from the 
L orbital to the L+1 orbital and vice versa. The other off-diagonal term gives rise to Hund's coupling between the singly-occupied 
L and L+1 orbitals on the same molecule \cite{Dutta14a}. The magnitudes of $U_{L,L+1}^{(d),\sigma,-\sigma}$ 
and $U_{L,L+1}^{(o),\sigma,-\sigma}$
are equal in the absence of long range e-e interactions. However, in the presence 
long range e-e interactions it is found that $|U_{L,L+1}^{(d),\sigma,-\sigma}|$ $>$ $|U_{L,L+1}^{(o),\sigma,-\sigma}|$ 
and occurs due to $\sum_{i,j;j\ne i} V_{i,j}\chi^{2}_{\mu,i,L}\chi_{\mu,j,L+1}^{2}$ being negative. Although the energies of the
L and L+1 orbitals, namely, $E_{L}$ and $E_{L+1}$, get renormalized, $\Delta_{L,L+1}$ does not change. Note that in 
Eq.~\ref{H_LL+1} the density-dependent hopping term (last term in Eq.~\ref{reduced2}) is absent. 
This is because density-dependent hoppings are important only when orbitals of same symmetry are involved. 
Thus for phenanthrene with FMOs of different symmetries, this term does not contribute. However, for 
phenacenes like picene, where each molecule can behave as a three-orbital system due to $\Delta_{L,L+1}$ = $\Delta_{L+1,L+2}$, 
this term is expected to have a non-zero contribution; $\Delta_{L+1,L+2}$ is the single-particle gap between the LUMO+1 LUMO+2
levels. 
\\

\bibliography{review}

\begin{thebibliography}{50}
\expandafter\ifx\csname natexlab\endcsname\relax\def\natexlab#1{#1}\fi
\expandafter\ifx\csname bibnamefont\endcsname\relax
  \def\bibnamefont#1{#1}\fi
\expandafter\ifx\csname bibfnamefont\endcsname\relax
  \def\bibfnamefont#1{#1}\fi
\expandafter\ifx\csname citenamefont\endcsname\relax
  \def\citenamefont#1{#1}\fi
\expandafter\ifx\csname url\endcsname\relax
  \def\url#1{\texttt{#1}}\fi
\expandafter\ifx\csname urlprefix\endcsname\relax\def\urlprefix{URL }\fi
\providecommand{\bibinfo}[2]{#2}
\providecommand{\eprint}[2][]{\url{#2}}

\bibitem[{\citenamefont{Tokura and Nagaosa}(2000)}]{Tokura00a}
\bibinfo{author}{\bibfnamefont{Y.}~\bibnamefont{Tokura}} \bibnamefont{and}
  \bibinfo{author}{\bibfnamefont{N.}~\bibnamefont{Nagaosa}},
  \bibinfo{journal}{Science} \textbf{\bibinfo{volume}{288}},
  \bibinfo{pages}{462} (\bibinfo{year}{2000}).

\bibitem[{\citenamefont{Imada et~al.}(1998)\citenamefont{Imada, Fujimori, and
  Tokura}}]{Imada98a}
\bibinfo{author}{\bibfnamefont{M.}~\bibnamefont{Imada}},
  \bibinfo{author}{\bibfnamefont{A.}~\bibnamefont{Fujimori}}, \bibnamefont{and}
  \bibinfo{author}{\bibfnamefont{Y.}~\bibnamefont{Tokura}},
  \bibinfo{journal}{Rev. Mod. Phys.} \textbf{\bibinfo{volume}{70}},
  \bibinfo{pages}{1039} (\bibinfo{year}{1998}).

\bibitem[{\citenamefont{Mackenzie and Maeno}(2003)}]{Mackenzie03a}
\bibinfo{author}{\bibfnamefont{A.~P.} \bibnamefont{Mackenzie}}
  \bibnamefont{and} \bibinfo{author}{\bibfnamefont{Y.}~\bibnamefont{Maeno}},
  \bibinfo{journal}{Rev.\ Mod.\ Phys.} \textbf{\bibinfo{volume}{75}},
  \bibinfo{pages}{657} (\bibinfo{year}{2003}).

\bibitem[{\citenamefont{Nakatsuji and Maeno}(2000)}]{Nakatsuji00a}
\bibinfo{author}{\bibfnamefont{S.}~\bibnamefont{Nakatsuji}} \bibnamefont{and}
  \bibinfo{author}{\bibfnamefont{Y.}~\bibnamefont{Maeno}},
  \bibinfo{journal}{Phys.\ Rev.\ Lett.} \textbf{\bibinfo{volume}{84}},
  \bibinfo{pages}{2666} (\bibinfo{year}{2000}).

\bibitem[{\citenamefont{Khaliullin and Maekawa}(2000)}]{Khaliullin00a}
\bibinfo{author}{\bibfnamefont{G.}~\bibnamefont{Khaliullin}} \bibnamefont{and}
  \bibinfo{author}{\bibfnamefont{S.}~\bibnamefont{Maekawa}},
  \bibinfo{journal}{Phys.\ Rev.\ Lett.} \textbf{\bibinfo{volume}{85}},
  \bibinfo{pages}{3950} (\bibinfo{year}{2000}).

\bibitem[{\citenamefont{Khaliullin et~al.}(2001)\citenamefont{Khaliullin,
  Horsch, and Oles}}]{Khaliullin01a}
\bibinfo{author}{\bibfnamefont{G.}~\bibnamefont{Khaliullin}},
  \bibinfo{author}{\bibfnamefont{P.}~\bibnamefont{Horsch}}, \bibnamefont{and}
  \bibinfo{author}{\bibfnamefont{A.~M.} \bibnamefont{Oles}},
  \bibinfo{journal}{Phys.\ Rev.\ Lett.} \textbf{\bibinfo{volume}{86}},
  \bibinfo{pages}{3879} (\bibinfo{year}{2001}).

\bibitem[{\citenamefont{Ishihara et~al.}(2002)\citenamefont{Ishihara,
  Hatakeyama, and Maekawa}}]{Ishihara02a}
\bibinfo{author}{\bibfnamefont{S.}~\bibnamefont{Ishihara}},
  \bibinfo{author}{\bibfnamefont{T.}~\bibnamefont{Hatakeyama}},
  \bibnamefont{and} \bibinfo{author}{\bibfnamefont{S.}~\bibnamefont{Maekawa}},
  \bibinfo{journal}{Phys.\ Rev.\ B} \textbf{\bibinfo{volume}{65}},
  \bibinfo{pages}{64442} (\bibinfo{year}{2002}).

\bibitem[{\citenamefont{Maeno et~al.}(1994)\citenamefont{Maeno, Hashimoto,
  Yoshida, Nishizaki, Fujita, Bednorz, and Lichtenberg}}]{Maeno94a}
\bibinfo{author}{\bibfnamefont{Y.}~\bibnamefont{Maeno}},
  \bibinfo{author}{\bibfnamefont{H.}~\bibnamefont{Hashimoto}},
  \bibinfo{author}{\bibfnamefont{K.}~\bibnamefont{Yoshida}},
  \bibinfo{author}{\bibfnamefont{S.}~\bibnamefont{Nishizaki}},
  \bibinfo{author}{\bibfnamefont{T.}~\bibnamefont{Fujita}},
  \bibinfo{author}{\bibfnamefont{J.~G.} \bibnamefont{Bednorz}},
  \bibnamefont{and}
  \bibinfo{author}{\bibfnamefont{F.}~\bibnamefont{Lichtenberg}},
  \bibinfo{journal}{Nature} \textbf{\bibinfo{volume}{372}},
  \bibinfo{pages}{532} (\bibinfo{year}{1994}).

\bibitem[{\citenamefont{Ishiguro et~al.}(1998)\citenamefont{Ishiguro, Yamaji,
  and Saito}}]{Ishiguro}
\bibinfo{author}{\bibfnamefont{T.}~\bibnamefont{Ishiguro}},
  \bibinfo{author}{\bibfnamefont{K.}~\bibnamefont{Yamaji}}, \bibnamefont{and}
  \bibinfo{author}{\bibfnamefont{G.}~\bibnamefont{Saito}},
  \emph{\bibinfo{title}{Organic Superconductors}}
  (\bibinfo{publisher}{Springer-Verlag}, \bibinfo{address}{New York},
  \bibinfo{year}{1998}).

\bibitem[{\citenamefont{Gunnarsson}(1997)}]{Gunnarsson97a}
\bibinfo{author}{\bibfnamefont{O.}~\bibnamefont{Gunnarsson}},
  \bibinfo{journal}{Rev.\ Mod.\ Phys.} \textbf{\bibinfo{volume}{69}},
  \bibinfo{pages}{575} (\bibinfo{year}{1997}).

\bibitem[{\citenamefont{Iwasa and Takenobu}(2003)}]{Iwasa03a}
\bibinfo{author}{\bibfnamefont{Y.}~\bibnamefont{Iwasa}} \bibnamefont{and}
  \bibinfo{author}{\bibfnamefont{T.}~\bibnamefont{Takenobu}},
  \bibinfo{journal}{J. Phys.: Condens. Matter} \textbf{\bibinfo{volume}{15}},
  \bibinfo{pages}{R495} (\bibinfo{year}{2003}).

\bibitem[{\citenamefont{Capone et~al.}(2009)\citenamefont{Capone, Fabrizio,
  Castellani, and Tosatti}}]{Capone09a}
\bibinfo{author}{\bibfnamefont{M.}~\bibnamefont{Capone}},
  \bibinfo{author}{\bibfnamefont{M.}~\bibnamefont{Fabrizio}},
  \bibinfo{author}{\bibfnamefont{C.}~\bibnamefont{Castellani}},
  \bibnamefont{and} \bibinfo{author}{\bibfnamefont{E.}~\bibnamefont{Tosatti}},
  \bibinfo{journal}{Rev.\ Mod.\ Phys.} \textbf{\bibinfo{volume}{81}},
  \bibinfo{pages}{943} (\bibinfo{year}{2009}).

\bibitem[{\citenamefont{Mitsuhashi et~al.}(2010)}]{Mitsuhashi10a}
\bibinfo{author}{\bibfnamefont{R.}~\bibnamefont{Mitsuhashi}}
  \bibnamefont{et~al.}, \bibinfo{journal}{Nature}
  \textbf{\bibinfo{volume}{464}}, \bibinfo{pages}{76} (\bibinfo{year}{2010}).

\bibitem[{\citenamefont{Kubozono et~al.}(2011)}]{Kubozono11a}
\bibinfo{author}{\bibfnamefont{Y.}~\bibnamefont{Kubozono}}
  \bibnamefont{et~al.}, \bibinfo{journal}{Phys. Chem. Chem. Phys.}
  \textbf{\bibinfo{volume}{13}}, \bibinfo{pages}{16476} (\bibinfo{year}{2011}).

\bibitem[{\citenamefont{Wang et~al.}(2011{\natexlab{a}})}]{Wang11a}
\bibinfo{author}{\bibfnamefont{X.}~\bibnamefont{Wang}} \bibnamefont{et~al.},
  \bibinfo{journal}{Nature Communications} \textbf{\bibinfo{volume}{2}},
  \bibinfo{pages}{507} (\bibinfo{year}{2011}{\natexlab{a}}).

\bibitem[{\citenamefont{Wang et~al.}(2011{\natexlab{b}})}]{Wang11b}
\bibinfo{author}{\bibfnamefont{X.~F.} \bibnamefont{Wang}} \bibnamefont{et~al.},
  \bibinfo{journal}{Phys.\ Rev.\ B} \textbf{\bibinfo{volume}{84}},
  \bibinfo{pages}{214523} (\bibinfo{year}{2011}{\natexlab{b}}).

\bibitem[{\citenamefont{Wang et~al.}(2012)}]{Wang12a}
\bibinfo{author}{\bibfnamefont{X.~F.} \bibnamefont{Wang}} \bibnamefont{et~al.},
  \bibinfo{journal}{J. Phys.: Condens. Matter} \textbf{\bibinfo{volume}{24}},
  \bibinfo{pages}{345701} (\bibinfo{year}{2012}).

\bibitem[{\citenamefont{Artioli and Malavasi}(2014)}]{Artioli14a}
\bibinfo{author}{\bibfnamefont{G.~A.} \bibnamefont{Artioli}} \bibnamefont{and}
  \bibinfo{author}{\bibfnamefont{L.}~\bibnamefont{Malavasi}},
  \bibinfo{journal}{J. Mater. Chem. C} \textbf{\bibinfo{volume}{2}},
  \bibinfo{pages}{1577} (\bibinfo{year}{2014}).

\bibitem[{\citenamefont{Yano et~al.}(2014)\citenamefont{Yano, Endo, Hasegawa,
  Okada, Yamada, and Sasaki}}]{Yano14a}
\bibinfo{author}{\bibfnamefont{M.}~\bibnamefont{Yano}},
  \bibinfo{author}{\bibfnamefont{M.}~\bibnamefont{Endo}},
  \bibinfo{author}{\bibfnamefont{Y.}~\bibnamefont{Hasegawa}},
  \bibinfo{author}{\bibfnamefont{R.}~\bibnamefont{Okada}},
  \bibinfo{author}{\bibfnamefont{Y.}~\bibnamefont{Yamada}}, \bibnamefont{and}
  \bibinfo{author}{\bibfnamefont{M.}~\bibnamefont{Sasaki}},
  \bibinfo{journal}{J. Chem. Phys.} \textbf{\bibinfo{volume}{141}},
  \bibinfo{pages}{034708} (\bibinfo{year}{2014}).

\bibitem[{\citenamefont{Merino and McKenzie}(2001)}]{Merino01a}
\bibinfo{author}{\bibfnamefont{J.}~\bibnamefont{Merino}} \bibnamefont{and}
  \bibinfo{author}{\bibfnamefont{R.~H.} \bibnamefont{McKenzie}},
  \bibinfo{journal}{Phys.\ Rev.\ Lett.} \textbf{\bibinfo{volume}{87}},
  \bibinfo{pages}{237002} (\bibinfo{year}{2001}).

\bibitem[{\citenamefont{Clay et~al.}(2012)\citenamefont{Clay, Dayal, Li, and
  Mazumdar}}]{Clay12b}
\bibinfo{author}{\bibfnamefont{R.~T.} \bibnamefont{Clay}},
  \bibinfo{author}{\bibfnamefont{S.}~\bibnamefont{Dayal}},
  \bibinfo{author}{\bibfnamefont{H.}~\bibnamefont{Li}}, \bibnamefont{and}
  \bibinfo{author}{\bibfnamefont{S.}~\bibnamefont{Mazumdar}},
  \bibinfo{journal}{Phys. Stat. Solidi} \textbf{\bibinfo{volume}{249}},
  \bibinfo{pages}{991} (\bibinfo{year}{2012}).

\bibitem[{\citenamefont{Dutta and Mazumdar}(2014)}]{Dutta14a}
\bibinfo{author}{\bibfnamefont{T.}~\bibnamefont{Dutta}} \bibnamefont{and}
  \bibinfo{author}{\bibfnamefont{S.}~\bibnamefont{Mazumdar}},
  \bibinfo{journal}{Phys.\ Rev.\ B} \textbf{\bibinfo{volume}{89}},
  \bibinfo{pages}{245129} (\bibinfo{year}{2014}).

\bibitem[{\citenamefont{Artioli et~al.}(2015)\citenamefont{Artioli, Hammerath,
  Mozzati, Carretta, Corana, Mannucci, Margadonnad, and Malavasi}}]{Artioli15a}
\bibinfo{author}{\bibfnamefont{G.~A.} \bibnamefont{Artioli}},
  \bibinfo{author}{\bibfnamefont{F.}~\bibnamefont{Hammerath}},
  \bibinfo{author}{\bibfnamefont{M.~C.} \bibnamefont{Mozzati}},
  \bibinfo{author}{\bibfnamefont{P.}~\bibnamefont{Carretta}},
  \bibinfo{author}{\bibfnamefont{F.}~\bibnamefont{Corana}},
  \bibinfo{author}{\bibfnamefont{B.}~\bibnamefont{Mannucci}},
  \bibinfo{author}{\bibfnamefont{S.}~\bibnamefont{Margadonnad}},
  \bibnamefont{and} \bibinfo{author}{\bibfnamefont{L.}~\bibnamefont{Malavasi}},
  \bibinfo{journal}{Chem. Commun.} \textbf{\bibinfo{volume}{51}},
  \bibinfo{pages}{1092} (\bibinfo{year}{2015}).

\bibitem[{\citenamefont{Ganin et~al.}(2008)}]{Ganin08a}
\bibinfo{author}{\bibfnamefont{A.~Y.} \bibnamefont{Ganin}}
  \bibnamefont{et~al.}, \bibinfo{journal}{Nature Materials}
  \textbf{\bibinfo{volume}{7}}, \bibinfo{pages}{367} (\bibinfo{year}{2008}).

\bibitem[{\citenamefont{Ganin et~al.}(2010)}]{Ganin10a}
\bibinfo{author}{\bibfnamefont{A.~Y.} \bibnamefont{Ganin}}
  \bibnamefont{et~al.}, \bibinfo{journal}{Nature}
  \textbf{\bibinfo{volume}{466}}, \bibinfo{pages}{221} (\bibinfo{year}{2010}).

\bibitem[{\citenamefont{Takabayashi et~al.}(2009)}]{Takabayashi09a}
\bibinfo{author}{\bibfnamefont{Y.}~\bibnamefont{Takabayashi}}
  \bibnamefont{et~al.}, \bibinfo{journal}{Science}
  \textbf{\bibinfo{volume}{323}}, \bibinfo{pages}{1585} (\bibinfo{year}{2009}).

\bibitem[{\citenamefont{Capone et~al.}(2002)\citenamefont{Capone, Fabrizio,
  Castellani, and Tosatti}}]{Capone02a}
\bibinfo{author}{\bibfnamefont{M.}~\bibnamefont{Capone}},
  \bibinfo{author}{\bibfnamefont{M.}~\bibnamefont{Fabrizio}},
  \bibinfo{author}{\bibfnamefont{C.}~\bibnamefont{Castellani}},
  \bibnamefont{and} \bibinfo{author}{\bibfnamefont{E.}~\bibnamefont{Tosatti}},
  \bibinfo{journal}{Science} \textbf{\bibinfo{volume}{296}},
  \bibinfo{pages}{2364} (\bibinfo{year}{2002}).

\bibitem[{\citenamefont{Nomura et~al.}(2016)\citenamefont{Nomura, Sakai,
  Capone, and Arita}}]{Nomura16a}
\bibinfo{author}{\bibfnamefont{Y.}~\bibnamefont{Nomura}},
  \bibinfo{author}{\bibfnamefont{S.}~\bibnamefont{Sakai}},
  \bibinfo{author}{\bibfnamefont{M.}~\bibnamefont{Capone}}, \bibnamefont{and}
  \bibinfo{author}{\bibfnamefont{R.}~\bibnamefont{Arita}}, \bibinfo{journal}{J.
  Phys.: Condens. Matter (Topical Reviews)} \textbf{\bibinfo{volume}{28}},
  \bibinfo{pages}{153001} (\bibinfo{year}{2016}).

\bibitem[{\citenamefont{Schmalian}(1998)}]{Schmalian98a}
\bibinfo{author}{\bibfnamefont{J.}~\bibnamefont{Schmalian}},
  \bibinfo{journal}{Phys.\ Rev.\ Lett.} \textbf{\bibinfo{volume}{81}},
  \bibinfo{pages}{4232} (\bibinfo{year}{1998}).

\bibitem[{\citenamefont{Vojta and Dagotto}(1999)}]{Vojta99a}
\bibinfo{author}{\bibfnamefont{M.}~\bibnamefont{Vojta}} \bibnamefont{and}
  \bibinfo{author}{\bibfnamefont{E.}~\bibnamefont{Dagotto}},
  \bibinfo{journal}{Phys.\ Rev.\ B} \textbf{\bibinfo{volume}{59}},
  \bibinfo{pages}{R713} (\bibinfo{year}{1999}).

\bibitem[{\citenamefont{Kyung and Tremblay}(2006)}]{Kyung06a}
\bibinfo{author}{\bibfnamefont{B.}~\bibnamefont{Kyung}} \bibnamefont{and}
  \bibinfo{author}{\bibfnamefont{A.~M.~S.} \bibnamefont{Tremblay}},
  \bibinfo{journal}{Phys.\ Rev.\ Lett.} \textbf{\bibinfo{volume}{97}},
  \bibinfo{pages}{046402} (\bibinfo{year}{2006}).

\bibitem[{\citenamefont{Powell and McKenzie}(2005)}]{Powell05a}
\bibinfo{author}{\bibfnamefont{B.~J.} \bibnamefont{Powell}} \bibnamefont{and}
  \bibinfo{author}{\bibfnamefont{R.~H.} \bibnamefont{McKenzie}},
  \bibinfo{journal}{Phys.\ Rev.\ Lett.} \textbf{\bibinfo{volume}{94}},
  \bibinfo{pages}{047004} (\bibinfo{year}{2005}).

\bibitem[{\citenamefont{Powell and McKenzie}(2007)}]{Powell07a}
\bibinfo{author}{\bibfnamefont{B.~J.} \bibnamefont{Powell}} \bibnamefont{and}
  \bibinfo{author}{\bibfnamefont{R.~H.} \bibnamefont{McKenzie}},
  \bibinfo{journal}{Phys.\ Rev.\ Lett.} \textbf{\bibinfo{volume}{98}},
  \bibinfo{pages}{027005} (\bibinfo{year}{2007}).

\bibitem[{\citenamefont{Clay et~al.}(2008)\citenamefont{Clay, Li, and
  Mazumdar}}]{Clay08a}
\bibinfo{author}{\bibfnamefont{R.~T.} \bibnamefont{Clay}},
  \bibinfo{author}{\bibfnamefont{H.}~\bibnamefont{Li}}, \bibnamefont{and}
  \bibinfo{author}{\bibfnamefont{S.}~\bibnamefont{Mazumdar}},
  \bibinfo{journal}{Phys.\ Rev.\ Lett.} \textbf{\bibinfo{volume}{101}},
  \bibinfo{pages}{166403} (\bibinfo{year}{2008}).

\bibitem[{\citenamefont{Dayal et~al.}(2012)\citenamefont{Dayal, Clay, and
  Mazumdar}}]{Dayal12a}
\bibinfo{author}{\bibfnamefont{S.}~\bibnamefont{Dayal}},
  \bibinfo{author}{\bibfnamefont{R.~T.} \bibnamefont{Clay}}, \bibnamefont{and}
  \bibinfo{author}{\bibfnamefont{S.}~\bibnamefont{Mazumdar}},
  \bibinfo{journal}{Phys.\ Rev.\ B} \textbf{\bibinfo{volume}{85}},
  \bibinfo{pages}{165141} (\bibinfo{year}{2012}).

\bibitem[{\citenamefont{Tocchio et~al.}(2009)\citenamefont{Tocchio, Parola,
  Gros, and Becca}}]{Tocchio09a}
\bibinfo{author}{\bibfnamefont{L.~F.} \bibnamefont{Tocchio}},
  \bibinfo{author}{\bibfnamefont{A.}~\bibnamefont{Parola}},
  \bibinfo{author}{\bibfnamefont{C.}~\bibnamefont{Gros}}, \bibnamefont{and}
  \bibinfo{author}{\bibfnamefont{F.}~\bibnamefont{Becca}},
  \bibinfo{journal}{Phys.\ Rev.\ B} \textbf{\bibinfo{volume}{80}},
  \bibinfo{pages}{064419} (\bibinfo{year}{2009}).

\bibitem[{\citenamefont{Watanabe et~al.}(2008)\citenamefont{Watanabe, Yokoyama,
  Tanaka, and Inoue}}]{Watanabe08a}
\bibinfo{author}{\bibfnamefont{T.}~\bibnamefont{Watanabe}},
  \bibinfo{author}{\bibfnamefont{H.}~\bibnamefont{Yokoyama}},
  \bibinfo{author}{\bibfnamefont{Y.}~\bibnamefont{Tanaka}}, \bibnamefont{and}
  \bibinfo{author}{\bibfnamefont{J.}~\bibnamefont{Inoue}},
  \bibinfo{journal}{Phys.\ Rev.\ B} \textbf{\bibinfo{volume}{77}},
  \bibinfo{pages}{214505} (\bibinfo{year}{2008}).

\bibitem[{\citenamefont{Gomes et~al.}(2013)\citenamefont{Gomes, Clay, and
  Mazumdar}}]{Gomes13a}
\bibinfo{author}{\bibfnamefont{N.}~\bibnamefont{Gomes}},
  \bibinfo{author}{\bibfnamefont{R.~T.} \bibnamefont{Clay}}, \bibnamefont{and}
  \bibinfo{author}{\bibfnamefont{S.}~\bibnamefont{Mazumdar}},
  \bibinfo{journal}{J. Phys. Condens Matter} \textbf{\bibinfo{volume}{25}},
  \bibinfo{pages}{385603} (\bibinfo{year}{2013}).

\bibitem[{\citenamefont{Yanagisawa}(2013)}]{Yanagisawa13a}
\bibinfo{author}{\bibfnamefont{T.}~\bibnamefont{Yanagisawa}},
  \bibinfo{journal}{New J. Phys.} \textbf{\bibinfo{volume}{15}},
  \bibinfo{pages}{033012} (\bibinfo{year}{2013}).

\bibitem[{\citenamefont{Craciun et~al.}(2009)}]{Craciun09a}
\bibinfo{author}{\bibfnamefont{M.~F.} \bibnamefont{Craciun}}
  \bibnamefont{et~al.}, \bibinfo{journal}{Phys.\ Rev.\ B}
  \textbf{\bibinfo{volume}{79}}, \bibinfo{pages}{125116}
  (\bibinfo{year}{2009}).

\bibitem[{\citenamefont{Pariser and Parr}(1953)}]{PPP53a}
\bibinfo{author}{\bibfnamefont{R.}~\bibnamefont{Pariser}} \bibnamefont{and}
  \bibinfo{author}{\bibfnamefont{R.~G.} \bibnamefont{Parr}},
  \bibinfo{journal}{J. Chem. Phys} \textbf{\bibinfo{volume}{21}},
  \bibinfo{pages}{466} (\bibinfo{year}{1953}).

\bibitem[{\citenamefont{Pople}(1953)}]{PPP53b}
\bibinfo{author}{\bibfnamefont{J.~A.} \bibnamefont{Pople}},
  \bibinfo{journal}{Trans. Farad. Soc} \textbf{\bibinfo{volume}{49}},
  \bibinfo{pages}{1375} (\bibinfo{year}{1953}).

\bibitem[{\citenamefont{Fabbiani et~al.}(2006)\citenamefont{Fabbiani, Allan,
  Parsons, and Pulham}}]{IUCR}
\bibinfo{author}{\bibfnamefont{F.~P.~A.} \bibnamefont{Fabbiani}},
  \bibinfo{author}{\bibfnamefont{D.~R.} \bibnamefont{Allan}},
  \bibinfo{author}{\bibfnamefont{S.}~\bibnamefont{Parsons}}, \bibnamefont{and}
  \bibinfo{author}{\bibfnamefont{C.~R.} \bibnamefont{Pulham}},
  \bibinfo{journal}{Acta Cryst.} \textbf{\bibinfo{volume}{B62}},
  \bibinfo{pages}{826} (\bibinfo{year}{2006}).

\bibitem[{\citenamefont{Huang et~al.}(2012)\citenamefont{Huang, Zhang, and
  Lin}}]{Huang12a}
\bibinfo{author}{\bibfnamefont{Z.}~\bibnamefont{Huang}},
  \bibinfo{author}{\bibfnamefont{C.}~\bibnamefont{Zhang}}, \bibnamefont{and}
  \bibinfo{author}{\bibfnamefont{H.-Q.} \bibnamefont{Lin}},
  \bibinfo{journal}{Scientific Reports} \textbf{\bibinfo{volume}{2}},
  \bibinfo{pages}{922} (\bibinfo{year}{2012}).

\bibitem[{\citenamefont{Chandross et~al.}(1997)\citenamefont{Chandross,
  Mazumdar, Liess, Lane, Vardeny, Hamaguchi, and Yoshino}}]{Chandross97a}
\bibinfo{author}{\bibfnamefont{M.}~\bibnamefont{Chandross}},
  \bibinfo{author}{\bibfnamefont{S.}~\bibnamefont{Mazumdar}},
  \bibinfo{author}{\bibfnamefont{M.}~\bibnamefont{Liess}},
  \bibinfo{author}{\bibfnamefont{P.~A.} \bibnamefont{Lane}},
  \bibinfo{author}{\bibfnamefont{Z.~V.} \bibnamefont{Vardeny}},
  \bibinfo{author}{\bibfnamefont{M.}~\bibnamefont{Hamaguchi}},
  \bibnamefont{and} \bibinfo{author}{\bibfnamefont{K.}~\bibnamefont{Yoshino}},
  \bibinfo{journal}{Phys.\ Rev.\ B} \textbf{\bibinfo{volume}{55}},
  \bibinfo{pages}{1486} (\bibinfo{year}{1997}).

\bibitem[{\citenamefont{Ohno}(1997)}]{Ohno64a}
\bibinfo{author}{\bibfnamefont{K.}~\bibnamefont{Ohno}},
  \bibinfo{journal}{Theor. Chim. Acta} \textbf{\bibinfo{volume}{55}},
  \bibinfo{pages}{1486} (\bibinfo{year}{1997}).

\bibitem[{\citenamefont{Soos and Ramasesha}(1984)}]{Soos84a}
\bibinfo{author}{\bibfnamefont{Z.~G.} \bibnamefont{Soos}} \bibnamefont{and}
  \bibinfo{author}{\bibfnamefont{S.}~\bibnamefont{Ramasesha}},
  \bibinfo{journal}{Phys.\ Rev.\ B} \textbf{\bibinfo{volume}{29}},
  \bibinfo{pages}{5410} (\bibinfo{year}{1984}).

\bibitem[{\citenamefont{Ramasesha and Soos}(1984)}]{SR84}
\bibinfo{author}{\bibfnamefont{S.}~\bibnamefont{Ramasesha}} \bibnamefont{and}
  \bibinfo{author}{\bibfnamefont{Z.~G.} \bibnamefont{Soos}},
  \bibinfo{journal}{Int. J. Quant. Chem.} \textbf{\bibinfo{volume}{XXV}},
  \bibinfo{pages}{1003} (\bibinfo{year}{1984}).

\bibitem[{\citenamefont{Ramasesha}(1986)}]{SR86}
\bibinfo{author}{\bibfnamefont{S.}~\bibnamefont{Ramasesha}},
  \bibinfo{journal}{Chem. Phys. Lett.} \textbf{\bibinfo{volume}{130}},
  \bibinfo{pages}{522} (\bibinfo{year}{1986}).

\bibitem[{\citenamefont{Chandross et~al.}(1999)\citenamefont{Chandross, Shimoi,
  and Mazumdar}}]{Chandross99a}
\bibinfo{author}{\bibfnamefont{M.}~\bibnamefont{Chandross}},
  \bibinfo{author}{\bibfnamefont{Y.}~\bibnamefont{Shimoi}}, \bibnamefont{and}
  \bibinfo{author}{\bibfnamefont{S.}~\bibnamefont{Mazumdar}},
  \bibinfo{journal}{Phys.\ Rev.\ B} \textbf{\bibinfo{volume}{59}},
  \bibinfo{pages}{4822} (\bibinfo{year}{1999}).

\end{thebibliography}

\end{document}